\documentclass[journal, 10pt]{IEEEtran}
\usepackage{amsmath,graphicx}
\usepackage{pifont}
\newcommand{\xmark}{\ding{53}}%
\usepackage{amssymb}
\usepackage{bm}
\usepackage{multirow}
\usepackage{adjustbox}
\usepackage{tabularx}
\usepackage{tikz}
\usepackage{caption}
\usepackage{subcaption}
\usepackage{verbatim}
\usepackage{float}
\usepackage{commath}
\usepackage{url}
\usepackage{xcolor}
\usepackage{soul}
\usepackage{colortbl}
\usepackage{hyperref}
\newcommand{\ceil}[1]{\left\lceil #1 \right\rceil}

\begin{document}

\title{Self-attending RNN for Speech Enhancement to Improve Cross-corpus Generalization }

\author{Ashutosh~Pandey, ~\IEEEmembership{Student~Member,~IEEE}~and~DeLiang~Wang,~\IEEEmembership{Fellow,~IEEE}

\thanks{This research was supported in part by two NIDCD grants (R01DC012048 and R02DC015521) and the Ohio Supercomputer
Center.}
\thanks{A. Pandey is with the Department of Computer Science and Engineering, The Ohio State University, Columbus, OH 43210 USA (e-mail:
pandey.99@osu.edu).}
\thanks{D. L. Wang is with the Department of Computer Science and Engineering
and the Center for Cognitive and Brain Sciences, The Ohio State University,
Columbus, OH 43210 USA (e-mail: dwang@cse.ohio-state.edu)}\vspace{-0.5em}}

\maketitle

\begin{abstract}
Deep neural networks (DNNs) represent the mainstream methodology for supervised speech enhancement, primarily due to their capability to model complex functions using hierarchical representations. However, a recent study revealed that DNNs trained on a single corpus fail to generalize to untrained corpora, especially in low signal-to-noise ratio (SNR) conditions. Developing a noise, speaker, and corpus independent speech enhancement algorithm is essential for real-world applications. In this study, we propose a self-attending recurrent neural network, or attentive recurrent network (ARN), for time-domain speech enhancement to improve cross-corpus generalization. ARN comprises of recurrent neural networks (RNNs) augmented with self-attention blocks and feedforward blocks. We evaluate ARN on different corpora with nonstationary noises in low SNR conditions. Experimental results demonstrate that ARN substantially outperforms competitive approaches to time-domain speech enhancement, such as RNNs and dual-path ARNs. Additionally, we report an important finding that the two popular approaches to speech enhancement: complex spectral mapping and time-domain enhancement, obtain similar results for RNN and ARN with large-scale training. We also provide a challenging subset of the test set used in this study for evaluating future algorithms and facilitating direct comparisons.
 \end{abstract}


\IEEEpeerreviewmaketitle

\section{Introduction}
Background noise is unavoidable in the real world. It reduces the intelligibility and quality of a speech signal for human listeners. Additionally, it can severely degrade the performance of speech-based applications, such as automatic speech recognition, speaker identification, and hearing aids. Speech enhancement aims at removing or attenuating background noise from a noisy speech signal. It is used as a preprocessor in speech-based applications to improve their performance in noisy environments. Monaural speech enhancement, which is the task of speech enhancement from single microphone recordings, is considered an extremely challenging problem, especially in the presence of nonstationary noises in low signal-to-noise ratio (SNR) conditions. This study focuses on monaural speech enhancement in the time domain.

Traditional approaches to monaural speech enhancement include spectral subtraction, Wiener filtering and statistical model-based methods \cite{loizou2013speech}. In recent years, supervised approaches to speech enhancement using deep neural networks (DNNs) have become the mainstream methodology for speech enhancement \cite{wang2017supervised}, primarily due to their capability to learn complex relations from supervised data by using hierarchical representations.

Speech enhancement mainly uses time-frequency representations, such as short-time Fourier transform (STFT), for extracting input features and training targets. Training targets play an important role in DNN performance and can be either masking based or mapping based. Masking based targets, such as ideal ratio mask \cite{wang2014training} and phase sensitive mask \cite{erdogan2015phase}, are based on time-frequency (T-F) relations between the noisy and the clean speech, whereas mapping based targets, such as spectral magnitude and log-spectral magnitude are based on clean speech  \cite{lu2013speech, xu2015regression}. DNN is trained in a supervised way to estimate training targets from input features. During evaluation, the enhanced waveform is obtained by reconstructing a signal from the estimated training target.

Most of the popular approaches to speech enhancement aim at enhancing only the spectral magnitude and use unaltered noisy phase for time-domain reconstruction \cite{lu2013speech, xu2015regression, weninger2015speech, chen2016large, fu2016snr, park2016fully, chen2017long, tan2018gated, pandey2018adversarial}. This is primarily due to a belief that spectral phase is unimportant for speech enhancement, and it exhibits no T-F structure amenable to supervised learning \cite{williamson2016complex}. However, a relatively recent study has demonstrated that phase can play an important role in the quality of enhanced speech, especially in low SNR conditions \cite{paliwal2011importance}. As a result, researchers have started exploring ways to enhance both the spectral magnitude and the spectral phase. The first study in this regard was done by Williamson et al. \cite{williamson2016complex}, where the Cartesian representation of STFT in terms of real and imaginary parts was used instead of the widely used polar representation to propose complex ratio masking due to the T-F structure in the Cartesian representation. Complex ratio masking was further utilized in many studies, such as \cite{choi2019phase, hu2020dccrn, zhou2021complex}. Complex spectral mapping, a related approach for jointly enhancing the magnitude and the phase, aims at directly predicting the real and the imaginary part of the clean spectrogram from the noisy spectrogram \cite{fu2017complex, pandey2019exploring, tan2019learning, pandey2020learning}.  

On the other hand, time-domain speech enhancement aims at directly predicting the clean speech samples from the noisy speech samples, and in the process, magnitude and phase are jointly enhanced \cite{fu2017raw}, \cite{pascual2017segan, rethage2017wavenet, qian2017speech, fu2018end, pandey2019new, pandey2019tcnn}. It does not require computations associated with the conversion of a signal to and from the frequency domain, and feature extraction becomes an implicit part of supervised learning. 


The study in {\cite{fu2017raw}} proposed a fully convolutional neural network (CNN) for time-domain speech enhancement. Time-domain speech enhancement was further improved by using better processing blocks, such as dilated convolutions {\cite{rethage2017wavenet}}, {\cite{qian2017speech}}, {\cite{pandey2019tcnn}}, dense connections {\cite{pandey2020densely}}, self-attention {\cite{giri2019attention}}, {\cite{pandey2021dense}}, and dual-path recurrent neural networks (RNNs) {\cite{luo2020dual}}, {\cite{pandey2020dual}}.

Additionally, time-domain speech enhancement has benefited from better optimization methods, such as adversarial training {\cite{pascual2017segan}} and better loss functions, such as a loss incorporating the objective metric of short-time objective intelligibility (STOI) {\cite{taal2011algorithm}}, {\cite{fu2018end}} or spectral magnitudes {\cite{Pandey2018}}, {\cite{pandey2019new}}. The STOI-based loss in {\cite{fu2018end}} was able to improve STOI but was found to be suboptimal for objective quality metric, perceptual evaluation of speech quality (PESQ) {\cite{pascual2017segan}}, and segmental SNR. The spectral magnitude based loss {\cite{Pandey2018}}, {\cite{pandey2019new}}, on the other hand, was able to improve both STOI and PESQ but was suboptimal for scale-invariant SNR. In {\cite{pandey2021dense}}, the spectral magnitude based loss was found to exhibit an artifact in the enhanced audio, which was subsequently removed by using an improved loss called phase constrained magnitude (PCM). The PCM loss not only removed the artifact but also obtained consistent improvement for different metrics, such as STOI, PESQ, and SNR.

Recently, it has been revealed that DNNs trained for speech enhancement do not generalize to untrained corpora, especially in low SNR conditions \cite{pandey2020cross}. Even time-domain enhancement networks, such as auto-encoder convolutional neural network (AECNN) \cite{pandey2019new} and temporal convolutional neural network (TCNN) \cite{pandey2019tcnn}, that exhibit strong performance for untrained speakers from the training corpus, fail to generalize to speakers from untrained corpora.  It is revealed that the corpus channel unwillingly acquired due to recording conditions is one of the main culprits for performance degradation from trained to untrained corpora. Several techniques were proposed to improve cross-corpus generalization, such as channel normalization, a better training corpus, and a smaller frame shift \cite{pandey2020cross}. The proposed techniques obtain significant improvements on untrained corpora for an IRM-based long short-term memory (LSTM) recurrent neural network (RNN). This work was further extended to complex spectral mapping with improved cross-corpus generalization \cite{pandey2020learning}. An interesting finding in \cite{pandey2020learning} is that a sophisticated architecture for complex spectral mapping, gated convolutional neural network (GCRN), which obtains impressive performance on trained corpora, fails to generalize to untrained corpora. Further, simple LSTM RNNs with a smaller frame shift are found to be very helpful for cross-corpus generalization.

Self-attention is a widely utilized mechanism for sequence-to-sequence tasks, such as machine translation \cite{vaswani2017attention}, image generation \cite{zhang2019self} and ASR \cite{dong2018speech}. It was first introduced in  \cite{vaswani2017attention}, which obtained start-of-the-art performance for sequence-to-sequence tasks by using networks comprising self-attention blocks only. In self-attention, a given output in a sequence is computed using a subset of the input sequence that is helpful for the output prediction. In other words, an output is predicted by attending to a subset of the input for improving output prediction. Many recent studies \cite{giri2019attention, zhao2020monaural, kim2020t, koizumi2020speech},  {\cite{nicolson2020masked}, \cite{pandey2020dual, pandey2021dense}, {\cite{roy2021deeplpc}, \cite{zhou2021complex} have employed self-attention for speech enhancement and reported significant improvements. 

Nicolson et al. {\cite{nicolson2020masked}} developed a network similar to the encoder of the transformer network {\cite{vaswani2017attention}} for a priori SNR estimation. The estimated SNR was used with a minimum-mean square error (MMSE) log-spectral amplitude estimator for magnitude enhancement. In a subsequent study {\cite{roy2021deeplpc}}, a similar network was employed for predicting linear predictive coding (LPC) power spectra, which was utilized with an augmented Kalman filter for time-domain speech enhancement. Zhao et al. {\cite{zhao2020monaural}} used self-attention within a CNN for spectral mapping based magnitude enhancement for speech dereverberation.

Self-attention for complex ratio masking and complex spectral mapping has been studied for speech enhancement. A complex-valued transformer with Gaussian-weighted self-attention mechanism was proposed in {\cite{kim2020t}}. A speaker-aware network using self-attention was investigated in {\cite{koizumi2020speech}}, and a self-attention mechanism within a convolutional recurrent network was utilized in {\cite{zhou2021complex}}. 
 
 The first study to use self-attention for time-domain speech enhancement was reported in {\cite{giri2019attention}}, which proposed a self-attention mechanism within a 1-dimensional UNet {\cite{ronneberger2015u}}. Pandey et al. proposed to use self-attention within layers of a dense UNet, which comprised dense blocks within encoder and decoder layers. A recent study {\cite{pandey2020dual}} also investigated self-attention with a dual-path RNN for time-domain speech enhancement. However, we find that time-domain self-attending networks, such as the ones in \cite{pandey2021dense} and \cite{pandey2020dual}, obtain subpar performance on untrained corpora.

In this work, we propose a self-attending RNN, or attentive recurrent network (ARN), for time-domain speech enhancement to improve cross-corpus generalization. ARN comprises RNN augmented with a self-attention block and a feedforward block. The proposed ARN is motivated by observations such as RNNs with a smaller frame shift are helpful for cross-corpus generalization \cite{pandey2020cross, pandey2020learning}, and self-attention is a general mechanism effective for speech enhancement \cite{giri2019attention, zhao2020monaural, kim2020t, koizumi2020speech},  \cite{nicolson2020masked}, \cite{pandey2020dual, pandey2021dense}, \cite{roy2021deeplpc}, \cite{zhou2021complex}.  We employ an efficient attention mechanism proposed specifically for RNN \cite{merity2019single}, which results in reduced memory consumption, faster training, and similar or better performance than the widely used attention mechanism in \cite{vaswani2017attention}. 

We find that self-attention mechanism in ARN leads to substantial improvement on untrained corpora. Further, ARN outperforms existing approaches to speech enhancement in terms of cross-corpus generalization. Additionally, we compare complex spectral mapping and time-domain enhancement for RNN and ARN and find that complex spectral mapping and time-domain enhancement obtain statistically similar results when trained on a large corpus. 

We find a subset of our test set to be particularly challenging for improving objective intelligibility and quality scores. To stimulate progress, we make this test set available online for evaluating future algorithms and facilitating direct comparisons.   

The rest of the paper is organized as follows.  Section II describes time-domain speech enhancement. Section III presents the details of ARN building blocks and Section IV describes ARN architecture for time-domain speech enhancement. Experimental settings are given in Section IV, and results and comparisons are presented in Section V. Concluding remarks are given in Section VI. 

\section{Time-domain speech enhancement}
A noisy speech signal $\bm{x}$ is defined as the sum of a clean speech signal $\bm{s}$ and a noise signal $\bm{n}$
\begin{equation}
\bm{x} = \bm{s} + \bm{n}
\end{equation}
 $\{\bm{x}, \bm{s},  \bm{n} \} \in \mathbb{R}^{M \times 1}$, and $M$ is the number of samples in the speech signal. A speech enhancement algorithm aims at obtaining a close estimate, $\bm{\hat{s}}$, of $\bm{s}$ given $\bm{x}$.

The goal of a time-domain speech enhancement algorithm is to compute $\bm{\hat{s}}$ directly from $\bm{x}$ instead of using a T-F representation of $\bm{x}$. Time-domain speech enhancement using a DNN can be formulated as
\begin{equation}
\bm{\hat{s}} = f_{\theta}(\bm{x})
\end{equation}
where $ f_{\theta}$ denotes a function represented by a DNN parametrized by $\theta$. 

\subsection{Frame-Level Processing}
Generally, a speech enhancement algorithm is designed to process frames of a speech signal. Given a noisy signal $\bm{x}$, it is first chunked into overlapping frames which is then processed at frame-level by a speech enhancement model. Let $\bm{X} \in \mathbb{R}^{T \times L} $ denote the matrix containing frames of signal $\bm{x}$ and $\bm{x}_{t} \in \mathbb{R}^{L \times 1}$ the $t^{th}$ frame.  $\bm{x}_{t}$ is defined as
\begin{equation}
x_{t}[k] = x[(t-1)\cdot J + k], \  k = 0, \cdots, L-1
\end{equation}
where $T$ is the number of frames, $L$ is the frame length, and $J$ is the frame shift. $T$ is given by $\ceil{\frac{M}{J}}$, where $\ceil{\ }$ denotes the ceiling function. $\bm{x}$ is padded with zeros if $M$ is not divisible by $J$. Frame-level processing using a DNN can be defined as
\begin{equation}
\label{eq_dnn_frame}
\widehat{\bm{s}}_{t} = f_{\bm{\theta}}(\bm{x}_{t-T_{1}}, \cdots, \bm{x}_{t-1}, \bm{x}_{t}, \bm{x}_{t+1}, \cdots, \bm{x}_{t+T_{2}})
\end{equation}
where $\widehat{\bm{s}}_{t}$ is computed using $\bm{x}_{t}$, $T_{1}$ past frames, and $T_{2}$ future frames.

\subsection{Causal Speech Enhancement}
A frame-level speech enhancement algorithm is considered causal if the estimation of a given frame $\bm{\hat{s}}_{t}$ is computed using noisy frames at time instances less than or equal to $t$. For causal speech enhancement Eq. (4) is modified as
\begin{equation}
\widehat{\bm{s}}_{t} = f_{\bm{\theta}}(\bm{x}_{t-T_{1}}, \cdots, \bm{x}_{t-1}, \bm{x}_{t})
\end{equation}
where $\widehat{\bm{s}}_{t}$ is computed using $\bm{x}_{t}$ and $T_{1}$ past frames. 

Causality is a necessary requirement for real-time speech enhancement. Further, we observe that a causal algorithm exhibits greater degradation on untrained corpora compared to a corresponding non-causal algorithm. Therefore, we also develop and compare causal algorithms.

\section{Attentive Recurrent Network }

A block diagram of ARN is given in Fig. 1. The building blocks of ARN are layer normalization, RNN, self-attention block, and feedforward block. Next, we describe these building blocks one by one. 
\begin{figure}[!h]
\centering
\includegraphics[width=0.5\textwidth, keepaspectratio]{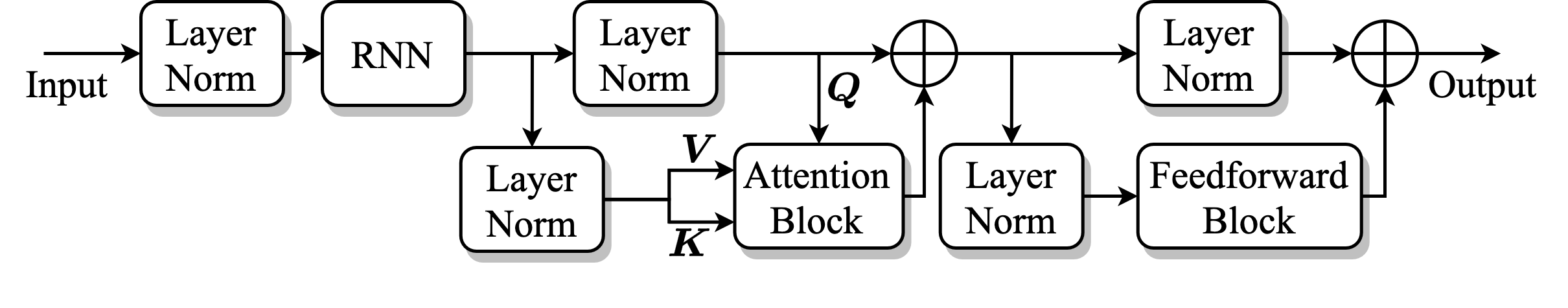}
\caption{A diagram of ARN. Layer Norm denotes a layer-normalization layer and $\bigoplus$ is an elementwise addition operator.}
\label{fig:rnnt}
\end{figure}

\subsection{Layer Normalization}
Layer normalization is a popular normalization technique used within DNNs to improve generalization and facilitate faster training \cite{ba2016layer}. It was proposed as an alternative to batch normalization \cite{ioffe2015batch}, which is found to be sensitive to training batch size. 

Let $\bm{X} \in \mathbb{R}^{T \times N}$ be a matrix and $\bm{x}_{t}$ be its $t^{th}$ row. We use the layer normalization defined as
\begin{equation}
\bm{x}_{t}^{norm} = \frac{\bm{x}_{t} - \mu_{x_{t}}}{\sqrt{\sigma^{2}_{x_{t}} + \epsilon}} \odot \bm{\gamma} + \bm{\beta}, \quad t = 1, \cdots, T
\end{equation}
where $ \mu_{x_{t}}$ and $\sigma_{x_{t}}^{2}$, respectively, are mean and variance of $\bm{x}_{t}$. Symbols $\bm{\gamma}$ and $ \bm{\beta}$ are trainable parameters of the same size as $\bm{x}_{t}$, $\odot$ denotes elementwise multiplication, and $\epsilon$ is a small positive constant used to avoid division by zero. 

\subsection {Recurrent Neural Network}
We use LSTM RNN in ARN. An illustrative diagram of an LSTM is shown in Fig. 2. Given an input vector sequence $\{ \bm{x}_{1}, \cdots, \bm{x}_{t-1}, \bm{x}_{t}, \bm{x}_{t+1}, \cdots, \bm{x}_{T}\}$, the hidden state at time $t$, $\bm{h}_{t}$, is computed as 
\begin{align}
\bm{i}_{t} &= \sigma (\bm{W}_{ix} \bm{x}_{t} + \bm{W}_{ih} \bm{h}_{t-1} + \bm{b}_{i}) \\
\bm{f}_{t} &= \sigma (\bm{W}_{fx} \bm{x}_{t} + \bm{W}_{fh} \bm{h}_{t-1} + \bm{b}_{f}) \\
\bm{g}_{t} &= \text{Tanh} (\bm{W}_{gx} \bm{x}_{t} + \bm{W}_{gh} \bm{h}_{t-1} + \bm{b}_{g}) \\
\bm{o}_{t} &= \sigma (\bm{W}_{ox} \bm{x}_{t} + \bm{W}_{oh} \bm{h}_{t-1} + \bm{b}_{o}) \\
\bm{c}_{t} &= \bm{f}_{t} \odot \bm{c}_{t-1} + \bm{i}_{t} \odot \bm{g}_{t} \\
\bm{h}_{t} &= \bm{o}_{t} \odot \text{Tanh}(\bm{c}_{t}) \\
\sigma (s) &= \frac{1}{1 + e^{-s}} \\
\text{Tanh}(s) &= \frac{e^{s} - e^{-s}}{e^{s} + e^{-s}}
\end{align}
where $\bm{x}_{t}$, $\bm{g}_{t}$,  and $\bm{c}_{t}$ respectively represent input, block input, and memory (cell) state at time $t$. In additions $\bm{i}_{t}$,  $\bm{f}_{t}$, and  $\bm{o}_{t}$ are gates known as input gate, forget gate and output gate, respectively. $\bm{W}$'s and $\bm{b}$'s denote trainable weights and biases.

\begin{figure}[!h]
\centering
\includegraphics[width=0.35\textwidth, keepaspectratio]{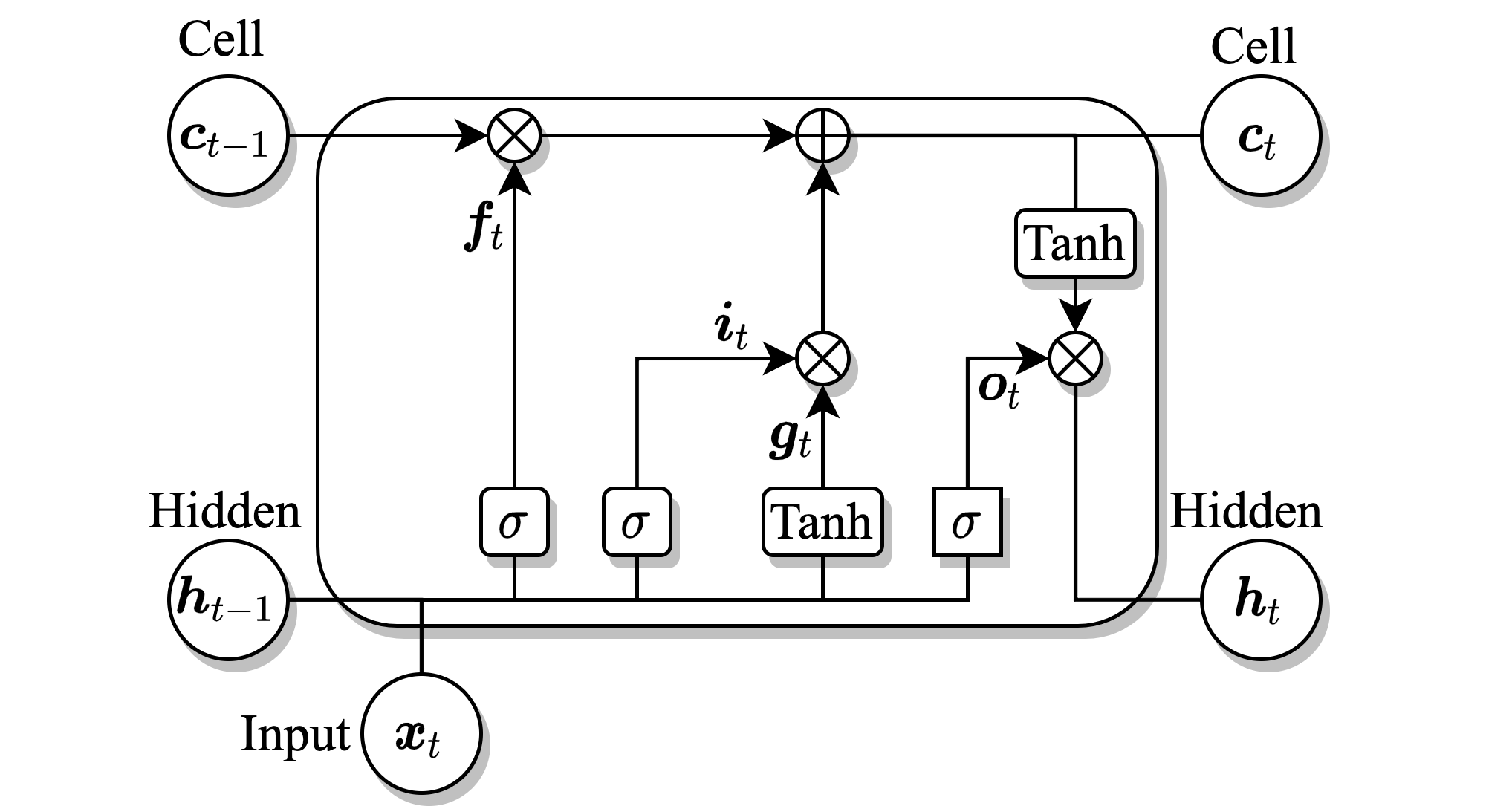}
\caption{A diagram of an LSTM with three gates. Symbol $\bigotimes$ denotes elementwise multiplication.}
\label{fig:rnnt}
\end{figure}

\subsection{Self-attention Block}
A general attention mechanism is defined using three components: key $\bm{K} \in \mathbb{R}^{T \times R}$, value $\bm{V} \in \mathbb{R}^{T \times S}$, and query $\bm{Q}$ $\in \mathbb{R}^{T \times R}$. First, correlation scores between pairs of rows from $\bm{Q}$ and $\bm{K}$, $\{ \bm{Q}_{i}, \bm{K}_{j}\}$, where $i, j \in \{1, \cdots , T \}$, are computed using the following equation.
\begin{equation}
 \bm{W} = \bm{Q}\bm{K}^{\mathbb{T}}
\end{equation}
where  $\bm{W} \in \mathbb{R}^{T \times T}$ and $\bm{K}^{\mathbb{T}}$ denotes the transpose of $\bm{K}$. The similarity scores in $\bm{W}$ are converted to probability values using the Softmax operation defined as 
\begin{equation}
\text{Softmax}(\bm{W})(i, j) = \frac{\exp^{W(i,j)}}{\sum_{j=0}^{T-1}\exp^{W(i, j)}}
\end{equation}

The final attention output is defined as the linear combination of the rows of $\bm{V}$ with weights in $\text{Softmax}(\bm{W})$.
\begin{equation}
\bm{A} = \text{Softmax}(\bm{W})\bm{V}
\end{equation}

In self-attention, $\bm{K}, \bm{V}$, and $\bm{Q}$ are computed from the same sequence. One of the approaches to self-attention is to use three linear projections of a given input, $\bm{X} \in \mathbb{R}^{T \times N}$, to obtain  $\bm{K}, \bm{V}$, and $\bm{Q}$, and then applying Eqs. (15)-(17) to obtain the attention output.

\begin{figure}[!h]
\centering
\includegraphics[width=0.35\textwidth, keepaspectratio]{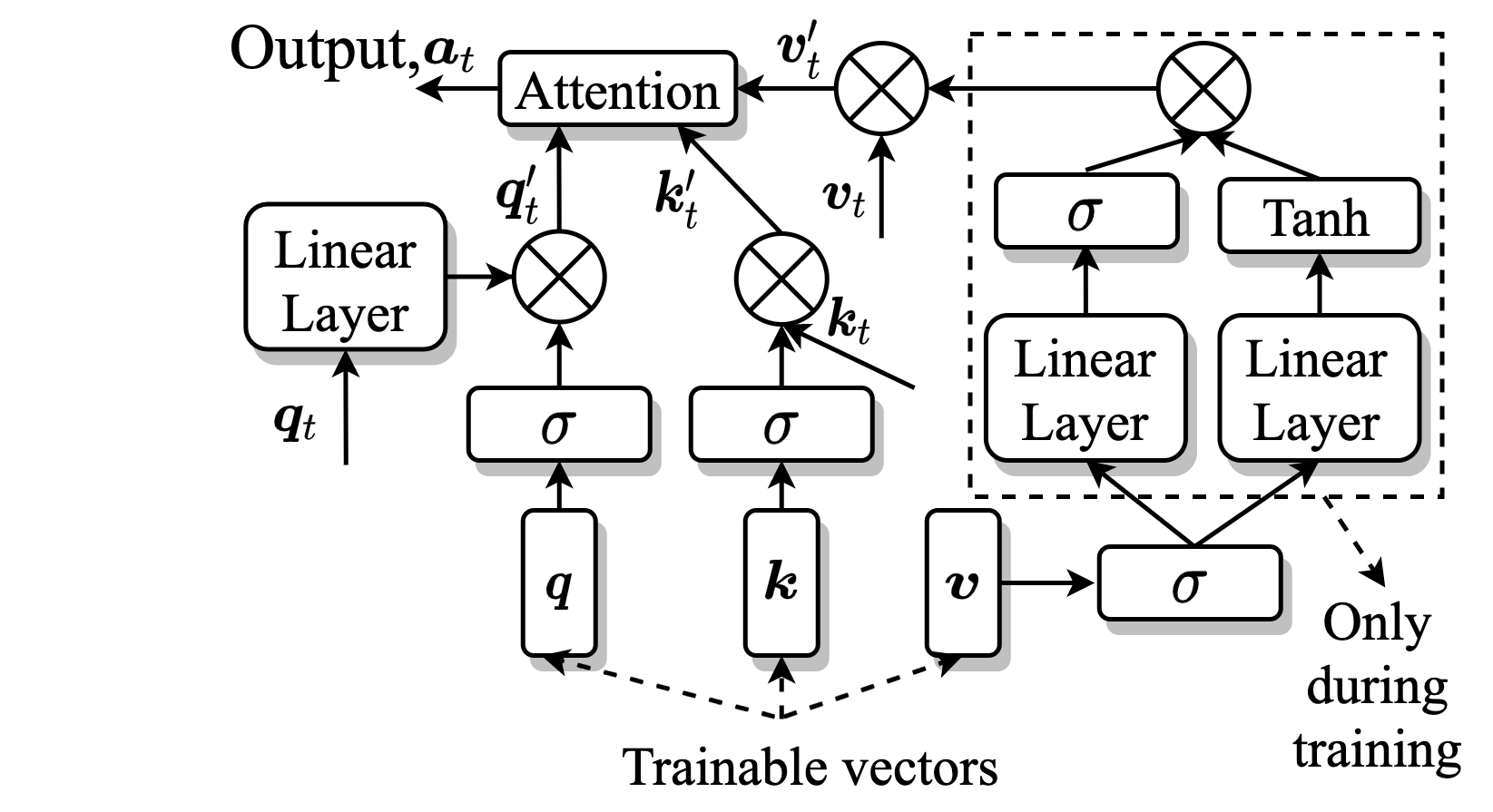}
\caption{Attention block in ARN. The inputs to the block are $\bm{q}_{t}, \bm{k}_{t}$, and $\bm{v}_{t}$ and the output from the block is $\bm{a}_{t}$.}
\label{fig:rnnt}
\end{figure}

A block diagram of the attention block in ARN is shown in Fig. 3. It comprises three trainable vectors $\{\bm{q}, \bm{k}$, $\bm{v} \} \in \mathbb{R}^{N \times 1}$ and its inputs are  $\{\bm{Q}, \bm{V}$, $\bm{K} \} \in \mathbb{R}^{T \times N}$. Let $\bm{q}_{t}, \bm{v}_{t}$, and  $\bm{k}_{t}$ denote the $t^{th}$ row in $\bm{Q}, \bm{V}$, and $\bm{K}$ respectively, and they are refined using gating mechanisms in the following equations.
\begin{align}
\bm{k}_{t}^{\prime} &= \bm{k}_{t} \odot \sigma(\bm{k})\\
\bm{q}_{t}^{\prime} &= \text{Lin}(\bm{q}_{t}) \odot  \sigma(\bm{q}) \\
\bm{v}_{t}^{\prime} &= \bm{v}_{t}\odot  [\sigma (\text{Lin}(\bm{v})) \odot \text{Tanh}(\text{Lin}(\bm{v}))]
\end{align}
where $\sigma $ is sigmoidal nonlinearity, and $\text{Lin()}$ is a linear layer. Note that $\sigma (\text{Lin}(\bm{v})) \odot \text{Tanh}(\text{Lin}(\bm{v}))$ represents a constant vector computed from $\bm{v}$. This operation is used during training only for better optimization of $\bm{v}$. For evaluation, we use its value from the best model at training completion.

 The final output of the attention block is computed as
 \begin{align}
 \bm{W}^{\prime} &= \frac{\bm{Q}^{\prime} \bm{K}^{\prime \mathbb{T}}}{\sqrt{N}} \\
 \bm{A} &= \text{Softmax}( \bm{W}^{\prime})\bm{V}^{\prime}
 \end{align}

\subsection{Feedforward Block}
The feedforward block in ARN is shown in Fig. 4. A given input of size $N$ is projected to size $4N$ using a linear layer, which is followed by Gaussian error linear unit (GELU) \cite{hendrycks2016gaussian} and a dropout layer \cite{srivastava2014dropout}. Finally, the output of size $4N$ is split into four vectors of size $N$, which are added together to get the final output.   

\begin{figure}[!h]
\centering
\includegraphics[width=0.38\textwidth, keepaspectratio]{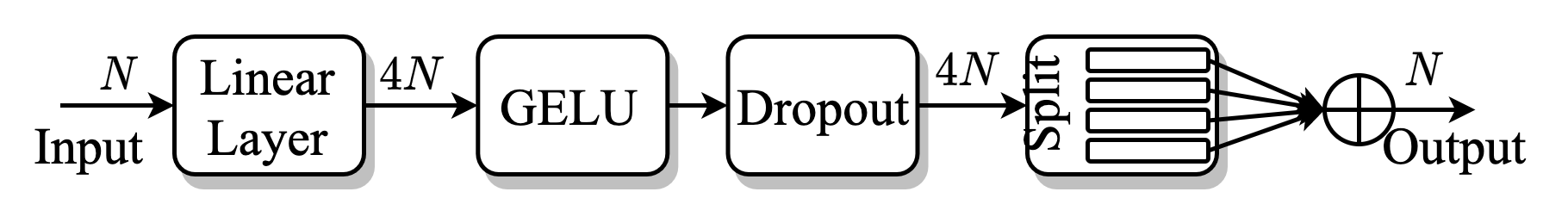}
\caption{Feedforward block in ARN.}
\label{fig:rnnt}
\end{figure}

With the building blocks described, we now present the processing flow of ARN shown in Fig. 1. The input to ARN is first normalized and then processed using an RNN. The output of the RNN is normalized using two parallel layer normalizations. The first stream is used as $\bm{Q}$ and the second stream is used as $\bm{K}$ and $\bm{V}$ for the following attention block. The output of the attention block is added to $\bm{Q}$ to form a residual connection. Again, the output is normalized using two parallel layer normalizations. The first stream is processed using a feedforward block and the second stream is added to the output of the feedforward block to form a residual connection. 

\section{ARN for Time-Domain Speech Enhancement}
The proposed ARN for time-domain speech enhancement is shown in Fig. 5. Given an input signal $\bm{x}$ with $M$ samples, it is first chunked into overlapping frames with a frame size of $L$ and frame shift of $J$ to obtain $T$ frames. Next, all the frames are projected to a representation of size $N$ using a linear layer, which is then processed using four ARNs. We use four-layered ARN as a simple extension of the four-layered RNN for complex spectral mapping in \cite{pandey2020learning}. A linear layer at the output projects the output of the last ARN to size $L$. Finally, overlap-and-add (OLA) is used to obtain the enhanced waveform.
\begin{figure}[!h]
\centering
\includegraphics[width=0.5\textwidth, keepaspectratio]{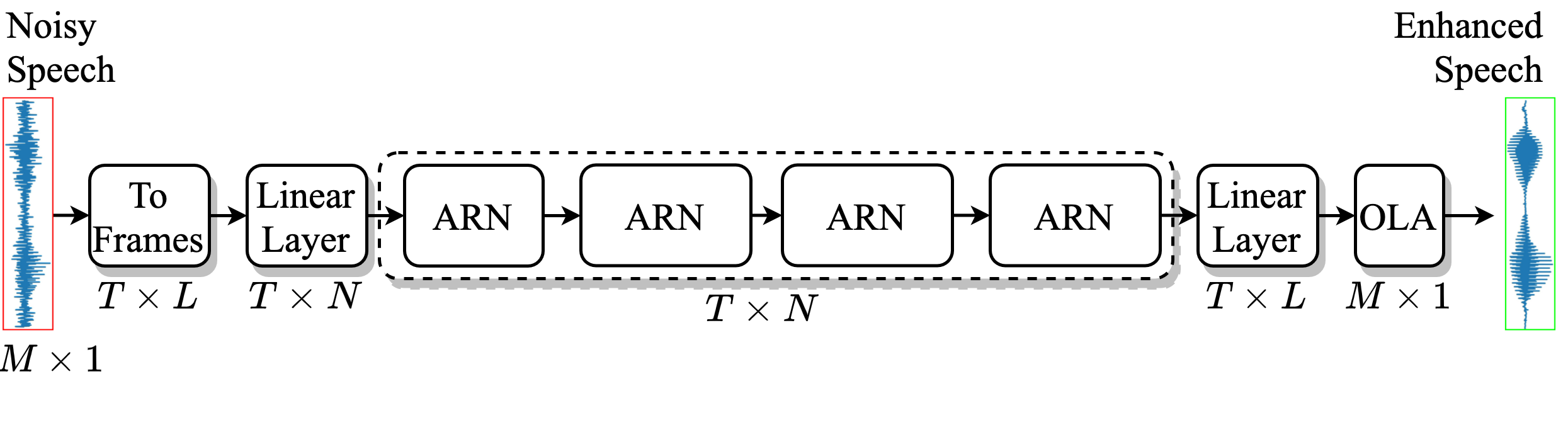}
\caption{The proposed ARN for time-domain speech enhancement.}
\label{fig:rnnt}
\end{figure}
\subsection{Non-causal Speech Enhancement}
For non-causal speech enhancement, we use BLSTM RNN inside ARN. A BLSTM comprises two LSTMs; a forward and a backward LSTM. The forward LSTM operates over the sequence in the original order, whereas the backward LSTM operates over the sequence in the reverse order. Let $\overrightarrow{\bm{X}}$ and  $\overleftarrow{\bm{X}}$ denote the sequence in the original and reverse order respectively. Then, we have
\begin{align}
\overrightarrow{\bm{x}}_{t} &= \bm{x}_{t} \\
\overleftarrow{\bm{x}}_{t} &= \bm{x}_{T-t}
\end{align}
The hidden state at time $t$ for a BLSTM is given as
\begin{equation}
\bm{h}_{t} = [\overrightarrow{\bm{h}}_{t}, \overleftarrow{\bm{h}}_{t}] = [\text{LSTM}_{f}(\overrightarrow{\bm{X}})_{t}, \text{LSTM}_{b}(\overleftarrow{\bm{X}})_{t}]
\end{equation}
where $[\bm{a}, \bm{b}]$ denotes a concatenation of vectors $\bm{a}$ and $\bm{b}$, and $\text{LSTM}_{f}$ and $\text{LSTM}_{b}$ represent the forward and the backward LSTM.

\subsection{Causal Speech Enhancement}
For causal speech enhancement, we use LSTM RNN and causal attention inside ARN. Causal attention is implemented by applying a mask to $\bm{W}^{\prime}$ where entries above the main diagonal are set to negative infinity so that the contribution from future frames in Eq. (22) becomes zero. The causal attention is defines as
\begin{equation}
\bm{A}_{causal} = \text{Softmax}(\text{Mask}(\bm{W}^{\prime}))\bm{V}^{\prime}
\end{equation}
where
\begin{equation}
	\text{Mask}(W^{\prime})(i, j) = \begin{cases}
	      W^{\prime}(i, j), & \text{if}\ i \leq j \\
	      -\infty, & \text{otherwise}
	    \end{cases}
\end{equation}

\section{Experimental Settings}
\subsection{Datasets}
We evaluate all the models in a speaker, noise, and corpus independent way. We use all utterances from the training set of LibriSpeech corpus \cite{panayotov2015librispeech} to generate training mixtures. It consists of around $280$K speech utterances of more than $2000$ speakers. LibriSpeech has been shown to be an effective corpus for cross-corpus generalization because it is recorded by many volunteers across the globe, and hence consists of utterances recorded in different acoustic conditions. Noisy training utterances are generated in an online fashion during training in the following way. For each sample in a given batch, we randomly sample a speech utterance, extract a random chunk of $4$ seconds from it, and add a random chunk of noise to it at a random SNR from $\{-5, -4, -3, -2, -1, 0 \}$ dB. The sampled speech is used unaltered if its duration is smaller than $4$ seconds. A set of $10000$ non-speech sounds from a sound effect library (\url{www.sound-ideas.com}) are used as the training noises. 

All the models are evaluated on three different corpora: WSJ-SI-$84$ (WSJ) \cite{paul1992design}, TIMIT \cite{garofolo1993darpa}, and IEEE \cite{rothauser1969ieee}, which are not used during training. We use utterances of one male speaker and one female speaker from IEEE to further categorize IEEE as IEEE Male and IEEE Female to show potential gender effects. The WSJ test set consists of $150$ utterances of $6$ different speakers. The TIMIT test set consists of $192$ utterances in the core test set. IEEE Male and IEEE Female each consists of $144$ randomly selected utterances. We generate noisy utterances using four different types of noises: babble, cafeteria, factory, and engine, none of which are used during training. Test utterances are generated at $6$ different SNRs of $-5$, $-2$, $0$, $2$, and $5$ dB. We find corpus fitting to be a severe issue for the difficult noises of babble and cafeteria, and at low SNRs of $-5$ dB and $-2$ dB. Therefore, for the sake of the brevity, we report results only for babble and cafeteria noises at SNRs of $-5$ dB and $-2$ dB. We observe similar performance trends for the other noises and SNR conditions. Note that our test set is the same as the one previously used in \cite{pandey2020cross} and \cite{pandey2020learning}. Babble and cafeteria noises are taken from an Auditec CD (available at \url{http://www.auditec.com}). Factory and engine noises are taken from the Noisex dataset \cite{varga1993assessment}. We use WSJ test utterances mixed with babble noise at the SNR of $-5$ dB as the validation set. 

We find our IEEE Male and IEEE Female test set to be relatively challenging in terms of improving the intelligibility and quality of unprocessed mixtures. In particular, IEEE utterances mixed with babble and cafeteria noises at SNRs of $-5$ dB and $-2$ dB are very difficult. Therefore, we provide online IEEE Male and IEEE Female utterances mixed with babble and cafeteria noises at SNRs of $-5$ dB and $-2$ dB as a useful test set for evaluating future algorithms and facilitating direct comparisons. It can be downloaded at \url{https://web.cse.ohio-state.edu/~wang.77/pnl/corpus/Pandey/NoisyIEEE.html}. 

In addition we investigate the proposed model for speech quality improvement in relatively high SNR conditions. We train ARN on the VCTK dataset {\cite{valentini2016investigating}} and compare it with a number of existing models evaluated on this task. The VCTK training set consists of utterances from $28$ speakers mixed with different noises at SNRs of $0$, $5$, $10$ and $15$ dB. We exclude two speakers (p274 and p282) from the training set to create a validation set. The test set comprises utterances from two unseen speakers (not in the training set) mixed with different noises at $2.5$, $7.5$, $12.5$, and $17.5$ dB. We store training speech and noises separately and dynamically mix them during training using random SNRs from $\{0, 5, 10, 15\}$ dB. The same SNR values are used as in the original training set. The dynamic mixing provides a measure of data augmentation, as similarly done in {\cite{defossez2020real}. 

We also evaluate our model on real recordings. We utilize the blind test set from the second deep noise suppression (DNS) challenge {\cite{reddy2021icassp}}, which consists of $650$ real recordings and $50$ synthetic mixtures. This test set is divided into five classes of English, non-English, tonal, singing, and emotional speech. To create a training set, we use speech and noises from the third DNS challenge {\cite{reddy2021interspeech}}. We remove low quality utterances from all classes using appropriate thresholds on provided parameter $T60\_WB$. We use room impulse responses (RIRs) with $T60\_WB$ between $0.3$ and $0.9$. The final training set consists of $347$K speech utterances, $65$K noises, and $47$K RIRs. We generate training mixtures dynamically by convolving a speech signal with a random RIR and adding a random noise segment. We add room reverberation with a probability of $0.5$. When using an RIR, the training target is set to be the clean speech convolved with the first $50$ ms of the RIR. We sample a SNR value uniformly from either $\{ -5, -4, \cdots, -1, 0 \}$ dB or from $\{ 1, 2, \cdots, 19, 20 \}$ dB with a probability of $0.5$. } 

\subsection{System Setup} 
All the utterances are resampled to $16$ kHz, and leading and trailing silences are removed from training utterances. Each noisy mixture is normalized using root mean square (RMS) normalization and the corresponding clean utterance is scaled accordingly to maintain an SNR. 

Parameter $N$ is set to $1024$, input frame size is set to $32$ ms for causal system and $16$ ms for non-causal system, and output frame size is set to $16$ ms. For ARN with BLSTM, $N = 1024$ results in a hidden state size of $512$ in both forward and backward LSTM. A dropout rate of $5\%$ is used in the feedforward block of ARN. We use the utterance level MSE (mean squared error) loss in the time domain for training on Librispeech and the PCM (phase constrained magnitude) loss {\cite{pandey2021dense}} for training on VCTK and DNS. The MSE loss is defined in the time domain as follows

\begin{equation}
L_{MSE}(\bm{s}, \bm{\hat{s}}) = \frac{1}{M}\sum_{k=0}^{M-1}(s[k] - \hat{s}[k])^2
\end{equation}

The PCM loss is defined in the T-F domain that measures the distance between clean and estimated magnitude spectrum of both speech and noise. It is defined using the following set of equations.
\begin{equation}
\bm{\widehat{n}} = \bm{x} - \bm{\widehat{s}}
\end{equation}
\begin{equation}
\begin{aligned}
L_{SM}(\bm{s}, \bm{\widehat{s}}) = \frac{1}{T\cdot F}\sum_{t=0}^{T-1}\sum_{f=0}^{F-1}&|(|S_{r}(t, f)| + |S_{i}(t,f)|)\\-&\ (|\widehat{S}_{r}(t,f)| + |\widehat{S}_{i}(t, f)|)|
\end{aligned}
\end{equation}
\begin{equation}
\begin{aligned}
L_{PCM}(\bm{s}, \bm{\widehat{s}}) &= \frac{1}{2}\cdot L_{SM}(\bm{s}, \bm{\widehat{s}}) + \frac{1}{2} \cdot L_{SM}(\bm{n}, \bm{\widehat{n}})
\end{aligned}
\end{equation}
where $\bm{S}$ and $\bm{\widehat{S}}$ respectively denote STFTs of $\bm{s}$ and $\bm{\widehat{s}}$, $T$ is the number of time frames, and $F$ is the number of frequency bins.

The Adam optimizer \cite{kingma2014adam} is used for training. A batch size of $32$ utterances is used on Librispeech and DNS and that of $8$ on VCTK. Models are trained for $100$ epochs on Librispeech, $84$ epochs on DNS, and $200$ epochs on VCTK. A constant learning rate of $0.0002$ is used for the first $33$ ($28$ for DNS) epochs, after which it is exponentially decayed using a scale that results in a learning rate of $0.00002$ in the final epoch. During training, we evaluate a given model on the validation set every two epochs, and the model parameters corresponding to the best SNR are chosen for evaluation.

We develop all the models in PyTorch \cite{paszke2017automatic} and exploit automatic mixed precision training to expedite training \cite{micikevicius2018mixed}. Two NVIDIA Volta V$100$ $32$GB GPUs are required to train ARN with a batch size of $32$ utterances. A given batch is equally distributed to two GPUs using PyTorch's DataParallel module. 

\subsection{Baseline Models}
We train five different models as the baselines for comparing corpus-independent models trained on Librispeech. First, we train a recently proposed deep complex convolutional recurrent network (DCCRN) \cite{hu2020dccrn}, which respectively won the first and the second place in real-time and non-real-time track of the first DNS challenge \cite{reddy2020interspeech}. DCCRN uses noisy complex spectrum as the input and the complex ideal ratio mask (cIRM) as the training target. Next, we train two RNN-based models; RNN-IRM \cite{pandey2020cross} and RNN-TCS \cite{pandey2020learning}. RNN-IRM uses log spectral magnitude as the input feature and the IRM as the training target. RNN-TCS uses noisy complex spectrum as the input feature and the target complex spectrum (TCS) as the training target. Finally, we train two recently proposed time-domain networks; dense convolutional network with self-attention (DCN) \cite{pandey2021dense} and dual-path ARN (DPARN) \cite{pandey2020dual}. Even though DCN and DPARN obtain good enhancement in the time domain, they have not been trained and evaluated in a corpus-independent way.

The ARN model trained on VCTK is compared with a number of existing methods that report performance on the same dataset (see Table III). The ARN model trained on the DNS challenge dataset is compared with a baseline noise suppression network (NSNet) provided with the third DNS challenge {\cite{reddy2021interspeech}}.

\subsection{Evaluation Metrics}

We use short-time objective intelligibility (STOI) \cite{taal2011algorithm} and narrow-band perceptual evaluation of speech quality (PESQ) \cite{rix2001perceptual} as evaluation metrics for comparing models trained on Librispeech. STOI has a typical range of $[0, 1]$, which roughly represents percent correct. PESQ has a range of $[-0.5, 4.5]$, where higher scores denote better speech quality.  Both metrics are commonly used for evaluating speech enhancement algorithms. For evaluating the models trained on VCTK, we use three metrics: composite scores {\cite{loizou2013speech}}, wide-band PESQ and STOI. Composite scores include three components: CSIG, CBAK and COVL, respectively measuring enhanced speech quality, noise removal and overall quality. For evaluation on the blind test set, we use a recently proposed non-intrusive quality metric, DNSMOS P.835, that highly correlates with subjective quality scores collected with the P.835 standard {\cite{reddy2021dnsmos}}. Similar to the composite scores, it has three components: DNSMOS-SIG, DNSMOS-BAK, and DNSMOS-OVL, respectively measuring enhanced speech quality, noise removal and overall quality.  

\section{Results and Discussions}
\begin{figure}[!t]
\centering
\includegraphics[width=0.48\textwidth]{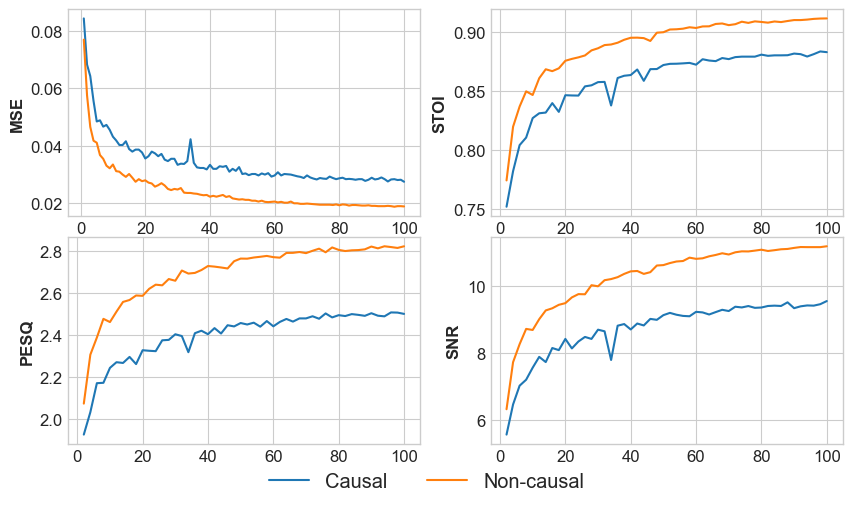}
\caption{Learning curves for ARN training on Librispeech with MSE Loss. We plot MSE loss, and STOI, PESQ and SNR scores on the validation set.}
\label{fig_model}
\end{figure}
\subsection{Learning Curves}
First, we plot performance curves of ARN training on Librispeech with MSE loss. Fig. 6 plots on the validation set the MSE loss every epoch, and STOI, PESQ, and SNR scores every other epoch. We can observe that, for both causal and non-causal models, training progresses smoothly and converges at the end with minimal improvements in last $10$ epochs.
\subsection{RNN vs ARN}
Now, we illustrate the effectiveness of self-attention for speech enhancement.  Fig. 7 plots average STOI and PESQ scores over babble and cafeteria noises for the four test corpora and at $2$ SNR conditions. The vertical bars at the top of the plots indicate $95\%$ confidence interval. We can observe that adding the proposed attention mechanism after each layer in RNN leads to significant improvements for all the test conditions. This suggests that self-attention is an effective mechanism for improving cross-corpus generalization of RNN-based speech enhancement. Note that improvements in cross-corpus generalization due to self-attention is not necessarily achieved in all architectures, as we find that DCN \cite{pandey2021dense}, a dense convolutional network with self-attention, fails to obtain similar improvements on untrained corpora (see Table I and Table II later).  

\begin{figure}[h]
\centering
\begin{subfigure}[t]{0.5\textwidth}
\centering
\includegraphics[width=0.96\textwidth]{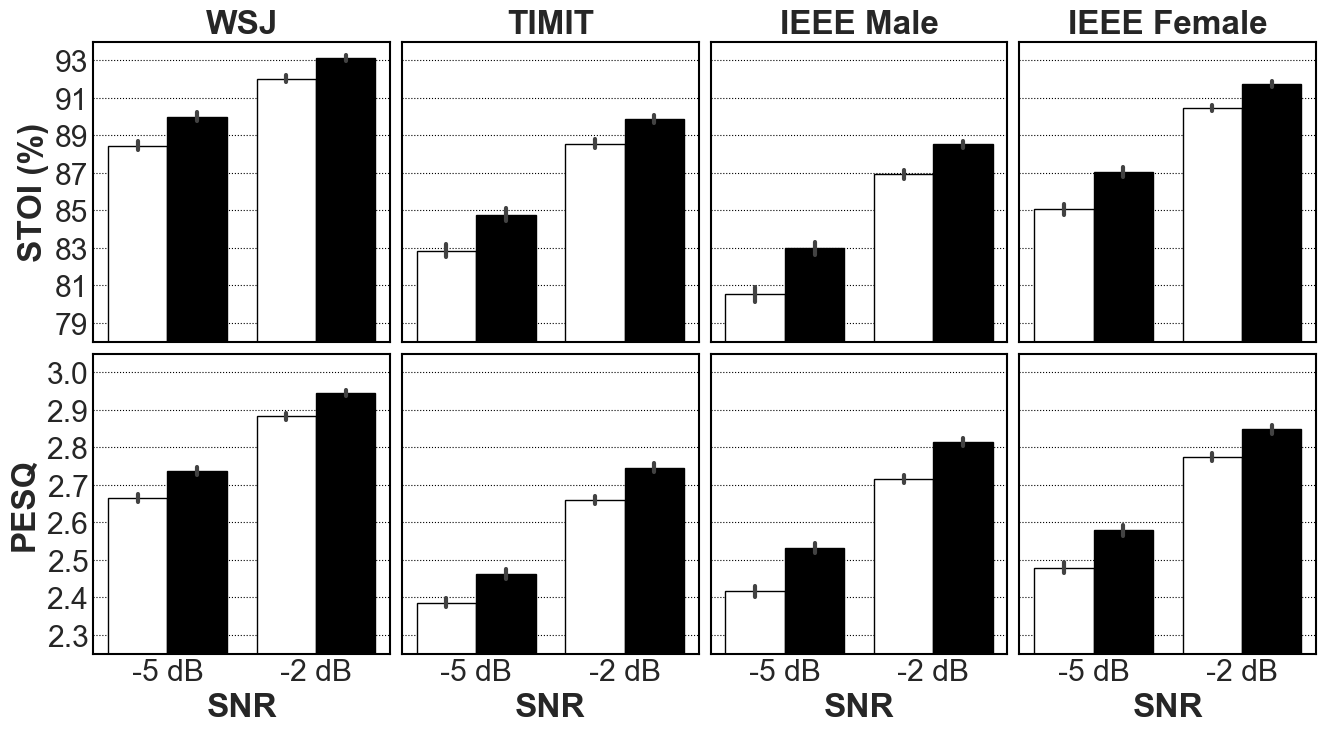}
\vspace{-0.8\baselineskip}
\caption{}
\end{subfigure}
\par\medskip
\begin{subfigure}[t]{0.5\textwidth}
\centering
\includegraphics[width=0.96\textwidth]{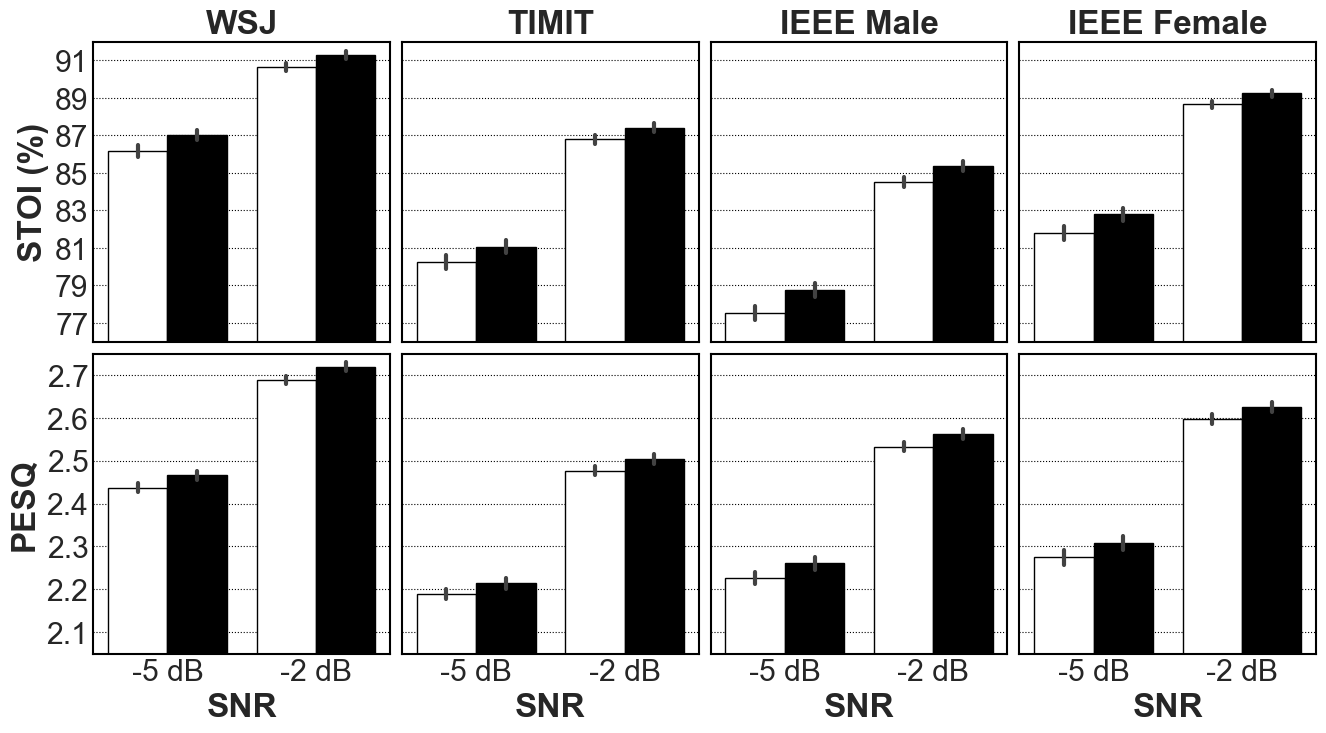}
\vspace{-0.8\baselineskip}
\caption{}
\end{subfigure}
\par\medskip
\begin{subfigure}[t]{0.5\textwidth}
\centering
\includegraphics[width=0.42\textwidth]{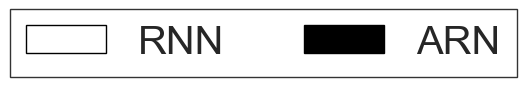}
\end{subfigure}
\caption{RNN comparisons with and without attention. a) Non-causal, b) causal.}
\label{fig_model}
\end{figure}

\subsection{Attention Mechanisms}
We compare two different attention mechanisms for ARN in causal and non-causal settings. Comparison results are plotted in Fig. 8. The first mechanism, denoted as A1, is the attention mechanism described in Section III-C. The second mechanism, denoted as A2, is borrowed from \cite{vaswani2017attention}, where we use one encoder layer without positional embeddings. We explore single-headed and 8-headed attention for this mechanism, which are respectively denoted as A2-1 and A2-8. We can observe that all the three attention mechanisms obtain statistically similar objective scores for both causal and non-causal speech enhancement. This suggests that even though self-attention is an effective technique for speech enhancement, changing the attention mechanism in ARN does not lead to statistically significant changes in the enhancement performance. 

\begin{figure}[t]
\centering
\begin{subfigure}[t]{0.5\textwidth}
\centering
\includegraphics[width=0.96\textwidth]{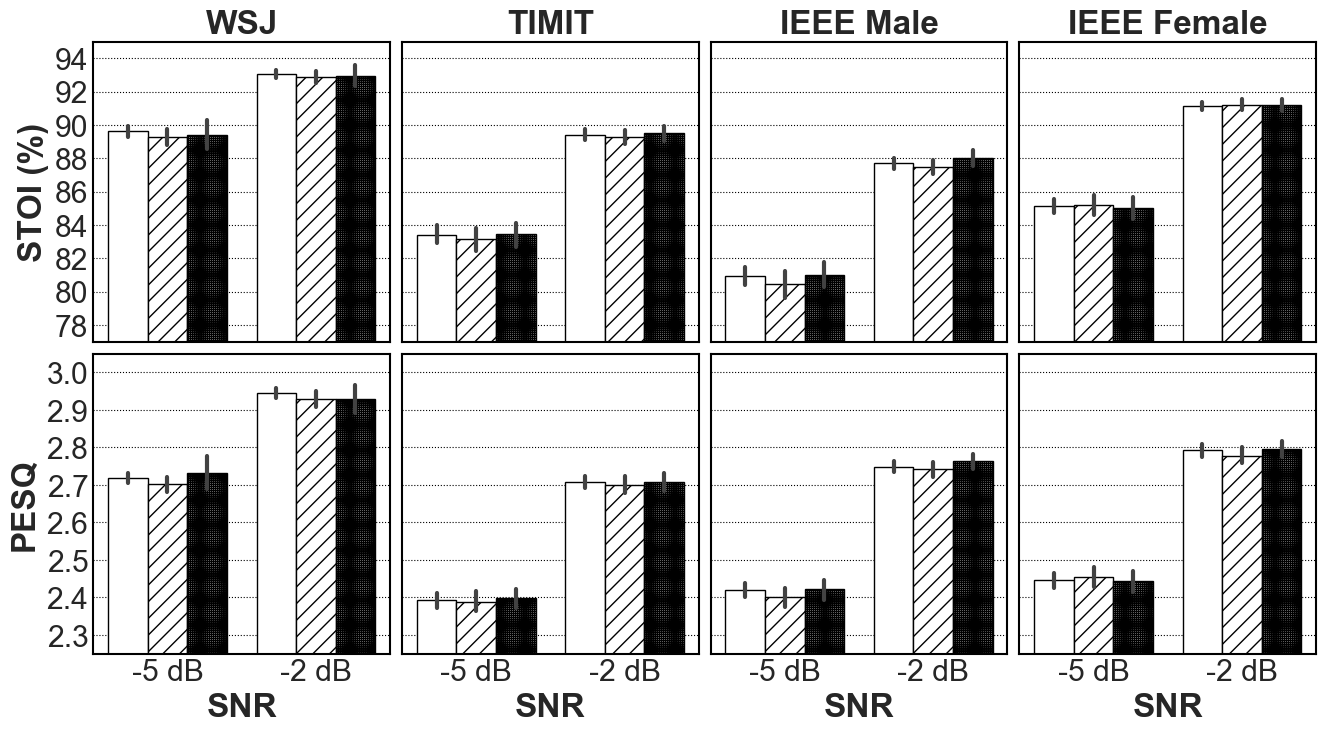}
\vspace{-0.8\baselineskip}
\caption{}
\end{subfigure}
\par\medskip
\begin{subfigure}[t]{0.5\textwidth}
\centering
\includegraphics[width=0.96\textwidth]{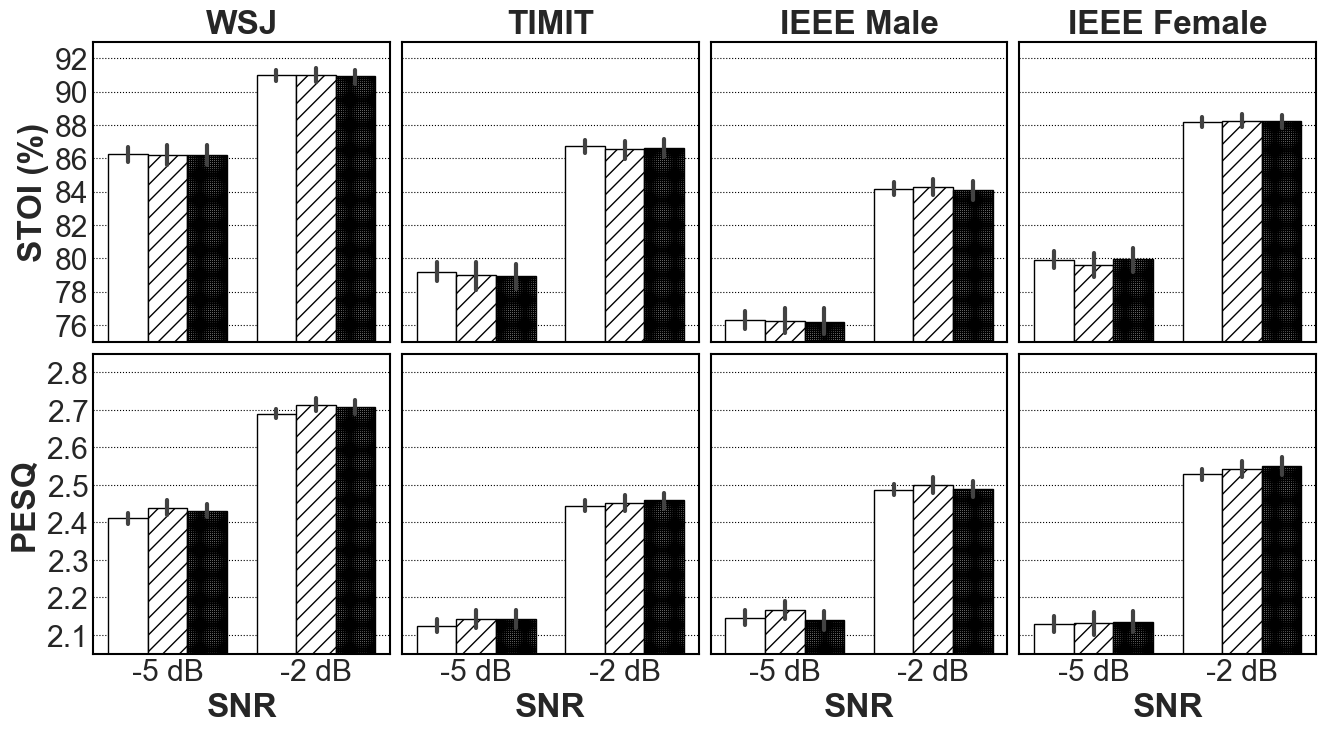}
\vspace{-0.8\baselineskip}
\caption{}
\end{subfigure}
\par\medskip
\begin{subfigure}[t]{0.5\textwidth}
\centering
\includegraphics[width=0.56\textwidth]{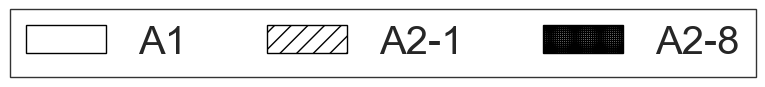}
\end{subfigure}
\caption{Comparisons of different attention mechanisms. a) Non-causal, b) causal.}
\label{fig_model}
\end{figure}

Next, in Fig. 9, we plot the number of parameters in ARN for the two attention mechanisms. We can see that there is a dramatic increase in the number of parameters when adding attention to an RNN-only network. However, the increase in the number of parameters due to A1 is roughly half of that due to A2. Also, we find A1 to be faster than A2 in both training and evaluation. As a result, we select A1 as the default attention mechanism in the remaining model comparisons.    

\begin{figure}[b]
\centering
\includegraphics[width=0.48\textwidth]{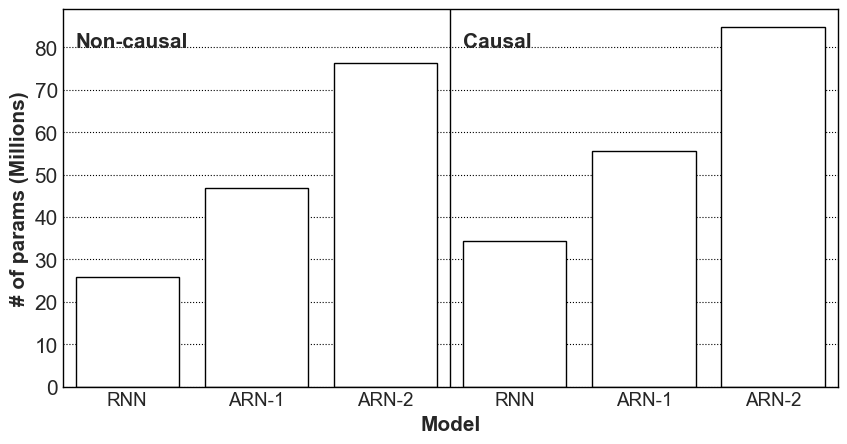}
\caption{Number of trainable parameters in ARN for different attention mechanisms.}
\label{fig_model}
\end{figure}

\subsection{Complex Spectral Mapping vs Time-domain Enhancement}
We evaluate RNN and ARN for both complex spectral mapping and time-domain speech enhancement. For complex spectral mapping, the input is the noisy STFT and the output is the estimated clean STFT. The real and the imaginary part of the STFT are concatenated together to obtain real-valued vectors. For time-domain enhancement, the input is the frames of the noisy speech and the output is the frames of the estimated clean speech. Average STOI and PESQ for two test noises and at two SNRs are plotted in Fig. 10. We can observe that time-domain enhancement is better than complex spectral mapping for most of the test cases; however, the performance difference is not statistically significant. Similar trends are observed with RNN and ARN for both causal and non-causal speech enhancement. This suggests that, with training on a large corpus such as LibriSpeech, complex spectral mapping and time-domain enhancement obtain similar results.

\begin{figure}[t]
\centering
\begin{subfigure}[t]{0.5\textwidth}
\centering
\includegraphics[width=0.96\textwidth]{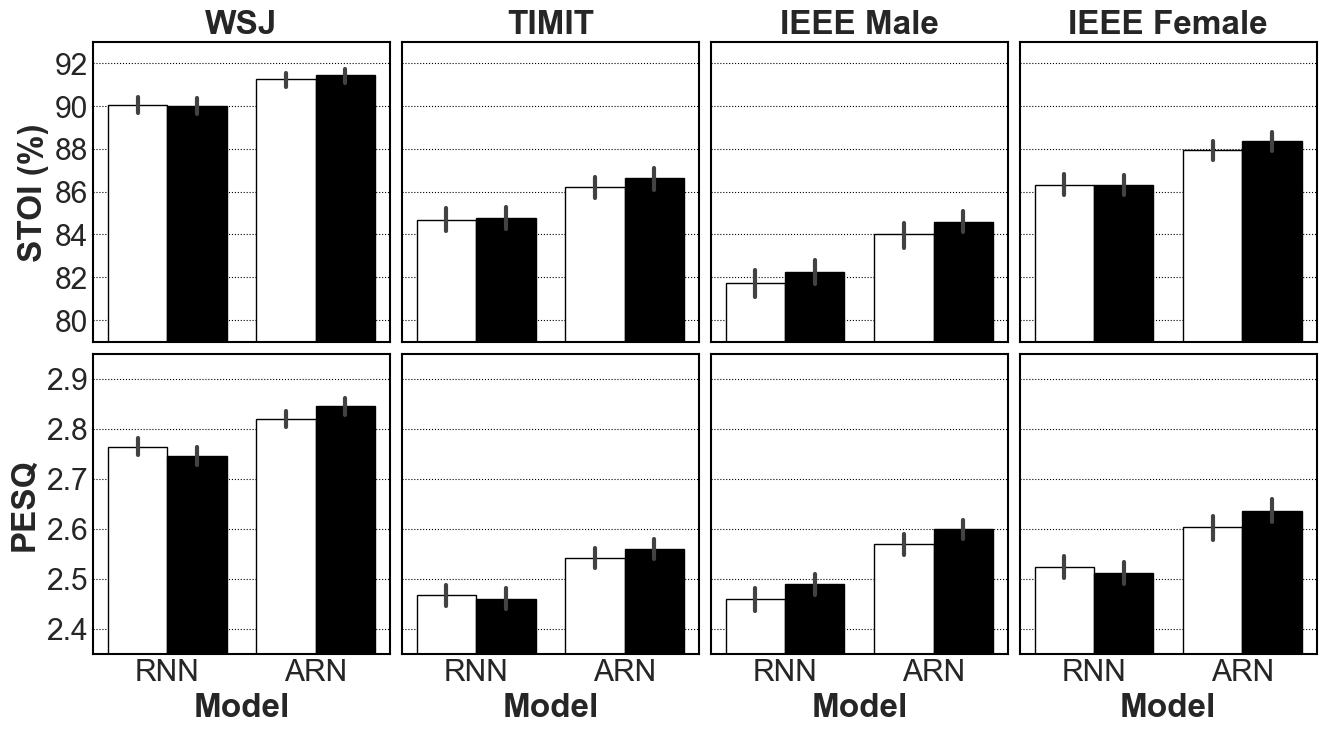}
\vspace{-0.8\baselineskip}
\caption{}
\end{subfigure}
\par\medskip
\begin{subfigure}[t]{0.5\textwidth}
\centering
\includegraphics[width=0.96\textwidth]{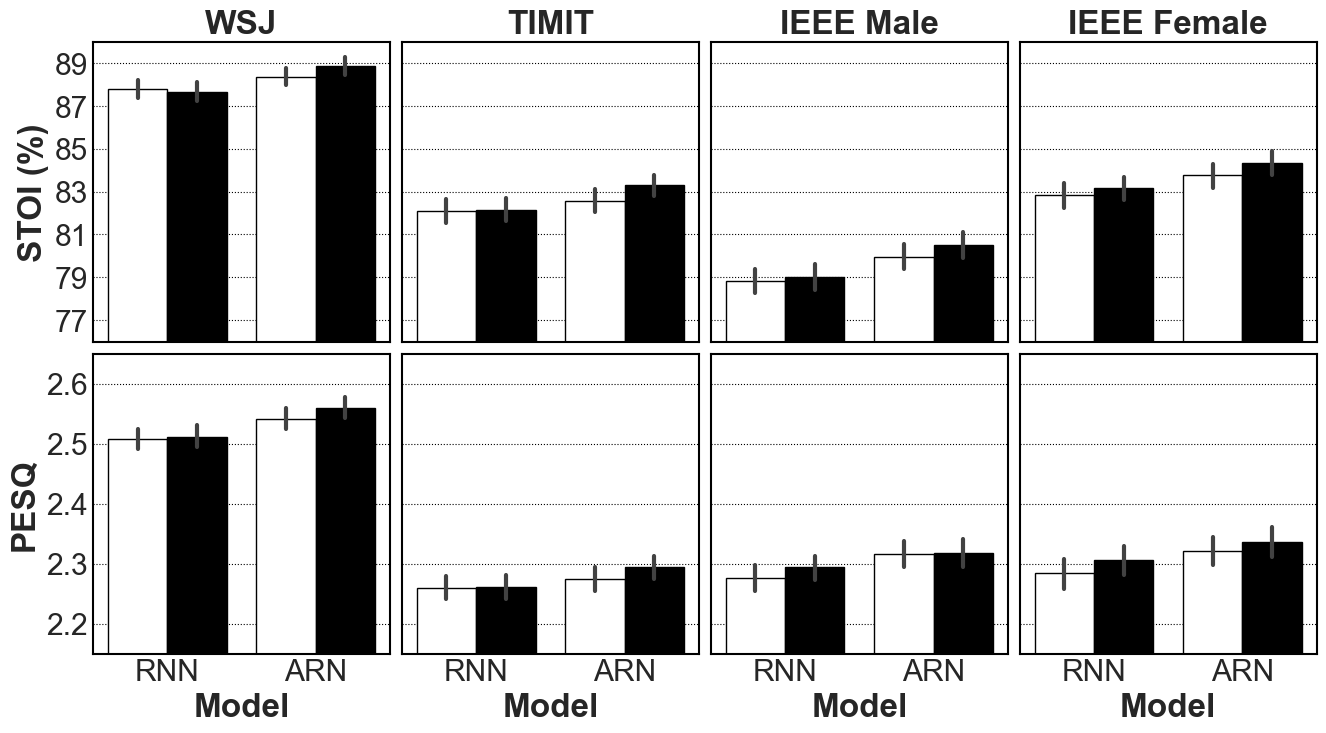}
\vspace{-0.8\baselineskip}
\caption{}
\end{subfigure}
\par\medskip
\begin{subfigure}[t]{0.5\textwidth}
\centering
\includegraphics[width=0.42\textwidth]{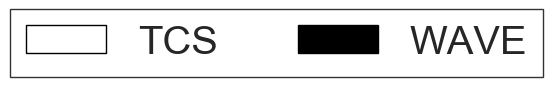}
\end{subfigure}
\caption{Comparing complex spectral mapping and time-domain enhancement for RNN and ARN. TCS stands for target complex spectrum and WAVE for waveform, which are respectively used as the training targets for complex spectral mapping and time-domain enhancement. a) Non-causal, b) causal.}
\label{fig_model}
\end{figure}

\subsection{Frame Shift}
Our previous studies in \cite{pandey2020cross} and \cite{pandey2020learning} suggest that a smaller frame shift leads to better speech enhancement on untrained corpora. As a result, a frame shift of $4$ ms is proposed for complex spectral mapping in \cite{pandey2020learning}. In this work, we are able to further decrease the frame shift from $4$ ms to $2$ ms with the help of automatic mixed precision training, which reduces memory consumption by half and improves training time significantly. Frame shifts of $4$ ms and $2$ ms are compared for RNN and ARN in Fig. 11. We observe that, except for the causal ARN at WSJ, a smaller frame shift leads to significant improvements for most of the test conditions. Similar performance trends are observed with RNN and ARN for both causal and non-causal enhancement. This further strengthens the argument that using a smaller frame shift is an effective technique for improving cross-corpus generalization. Note that it was reported in \cite{pandey2020learning} that for a gated convolutional recurrent network (GCRN) \cite{tan2019learning}, a smaller frame shift does not always lead to better cross-corpus generalization. It might be due to the fact that the receptive field of a convolutional neural network is constant, and as a result, reducing the frame shift leads to a reduction in the effective receptive field.  

\begin{figure}[t]
\centering
\begin{subfigure}[t]{0.5\textwidth}
\centering
\includegraphics[width=0.96\textwidth]{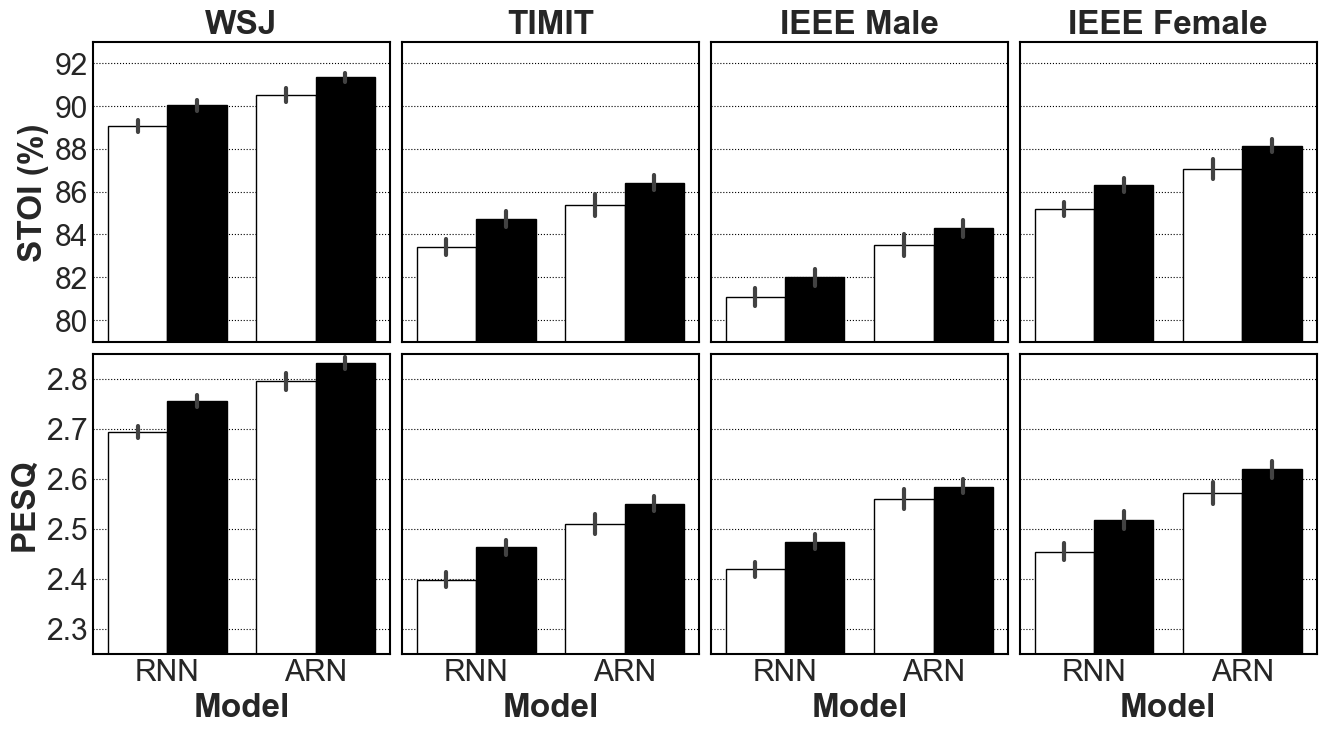}
\vspace{-0.8\baselineskip}
\caption{}
\end{subfigure}
\par\medskip
\begin{subfigure}[t]{0.5\textwidth}
\centering
\includegraphics[width=0.96\textwidth]{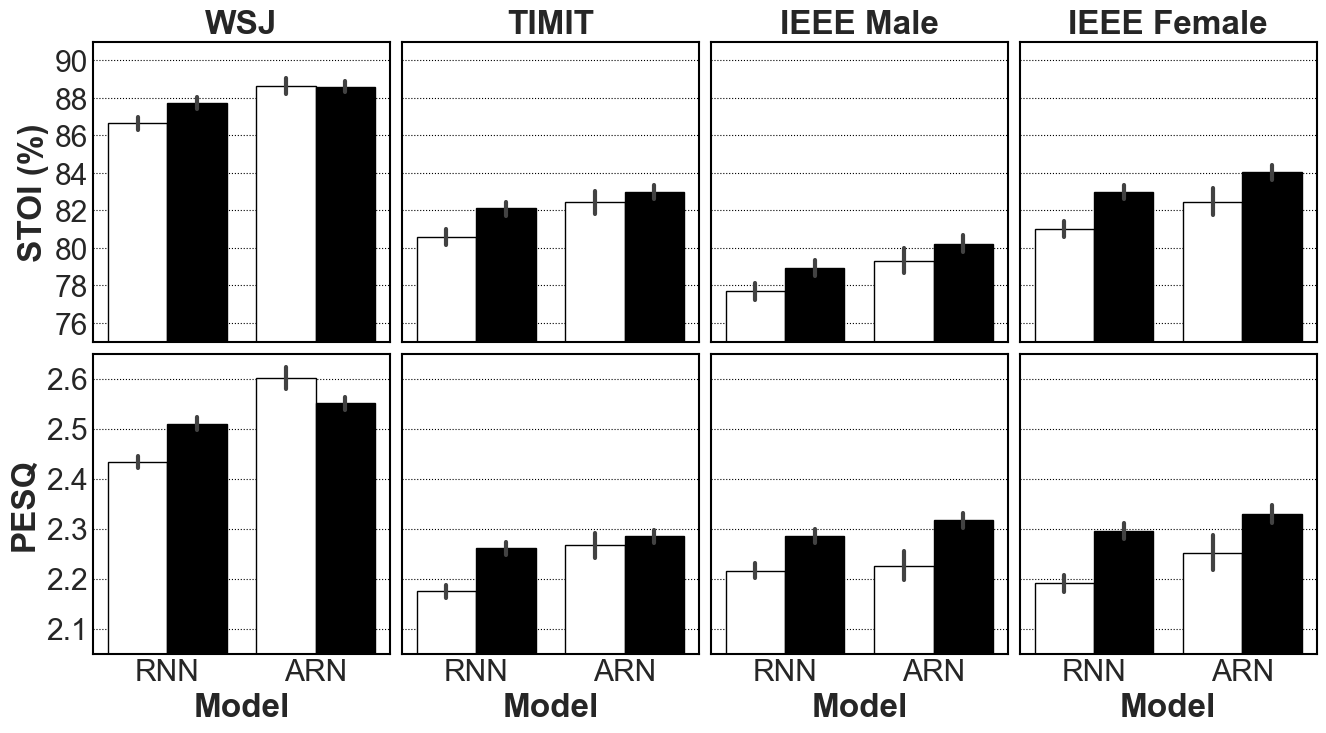}
\vspace{-0.8\baselineskip}
\caption{}
\end{subfigure}
\par\medskip
\begin{subfigure}[t]{0.5\textwidth}
\centering
\includegraphics[width=0.42\textwidth]{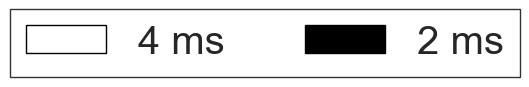}
\end{subfigure}
\caption{Effects of frame shifts for RNN and ARN. a) Non-causal, b) causal.}
\label{fig_model}
\end{figure}

\begin{table*}[t]
\centering
\caption{Comparing non-causal ARN with other non-causal approaches to speech enhancement.}
\label{tbl_ablation}
\centering
\begin{adjustbox}{width=0.8\textwidth}
\begin{tabular}{|c|c|cc|cc|cc|cc|cc|cc|cc|cc|}
\hline
\multicolumn{2}{|c|}{Test Noise} & \multicolumn{8}{c|}{Babble} & \multicolumn{8}{c|}{Cafeteria} \\
\hline
\multicolumn{2}{|c|}{Test Corpus} & \multicolumn{2}{c|}{WSJ} & \multicolumn{2}{c|}{TIMIT} & \multicolumn{2}{c|}{IEEE Male} & \multicolumn{2}{c|}{IEEE Female} & \multicolumn{2}{c|}{WSJ} & \multicolumn{2}{c|}{TIMIT} & \multicolumn{2}{c|}{IEEE Male} & \multicolumn{2}{c|}{IEEE Female} \\
\hline
\multicolumn{2}{|c|}{Test SNR} & -5 dB & -2 dB & -5 dB & -2 dB & -5 dB & -2 dB & -5 dB & -2 dB & -5 dB & -2 dB & -5 dB & -2 dB & -5 dB & -2 dB & -5 dB & -2 dB \\
\hline
\hline
\multirow{8}{*}{ \rotatebox{90}{STOI (\%) }} & Mixture & 58.6 & 65.5 & 54.0 & 60.9 & 55.0 & 62.3 & 55.5 & 62.9 & 57.4 & 64.5 & 53.1 & 60.1 & 54.8 & 60.9 & 55.1 & 62.0 \\
\cline{2-18}
& DCCRN & 82.5 & 89.0 & 73.1 & 82.5 & 68.3 & 81.3 & 72.5 & 84.6 & 81.4 & 87.6 & 74.8 & 82.6 & 72.0 & 80.3 & 77.4 & 86.0 \\
& RNN-IRM \cite{pandey2020cross} & 83.7 & 88.4 & 76.3 & 83.3 & 75.7 & 84.1 & 76.0 & 85.6 & 81.9 & 86.9 & 76.3 & 82.3 & 74.5 & 81.5 & 78.8 & 85.3 \\
& RNN-TCS \cite{pandey2020learning} & 88.1 & 92.2 & 79.3 & 87.5 & 76.7 & 85.8 & 80.0 & 89.2 & 85.8 & 90.3 & 80.4 & 86.6 & 77.3 & 84.1 & 82.6 & 88.7 \\
& DCN & 87.1 & 91.5 & 77.9 & 86.1 & 73.9 & 84.3 & 76.6 & 87.7 & 84.9 & 89.7 & 78.7 & 85.4 & 75.6 & 83.3 & 79.7 & 87.4 \\
& DPARN & 90.5 & 93.6 & 82.9 & 89.6 & 78.4 & 87.4 & 84.2 & 91.1 & 87.5 & 91.4 & 81.8 & 88.0 & 78.7 & 85.5 & 83.2 & 89.5 \\
& ARN & \textbf{91.1} & \textbf{94.1} & \textbf{84.5} & \textbf{90.6} & \textbf{82.3} & \textbf{88.9} & \textbf{85.6} & \textbf{92.0} & \textbf{88.3} & \textbf{92.1} & \textbf{82.7} & \textbf{88.6} & \textbf{80.6} & \textbf{86.6} & \textbf{85.3} & \textbf{90.5} \\

\hline
\hline
\multirow{8}{*}{  \rotatebox{90}{PESQ }} & Mixture & 1.54 & 1.69 & 1.46 & 1.63 & 1.45 & 1.63 & 1.12 & 1.32 & 1.44 & 1.64 & 1.33 & 1.52 & 1.37 & 1.54 & 1.01 & 1.20 \\
\cline{2-18}
& DCCRN & 2.31 & 2.65 & 1.99 & 2.38 & 1.86 & 2.33 & 1.79 & 2.33 & 2.32 & 2.61 & 2.12 & 2.39 & 2.08 & 2.40 & 2.14 & 2.50 \\
& RNN-IRM  & 2.51 & 2.82 & 2.27 & 2.60 & 2.15 & 2.54 & 2.00 & 2.51 & 2.49 & 2.76 & 2.31 & 2.57 & 2.21 & 2.51 & 2.22 & 2.57 \\
& RNN-TCS & 2.63 & 2.89 & 2.22 & 2.59 & 2.20 & 2.59 & 2.18 & 2.62 & 2.52 & 2.76 & 2.26 & 2.53 & 2.27 & 2.59 & 2.34 & 2.65 \\
& DCN & 2.56 & 2.85 & 2.14 & 2.50 & 2.09 & 2.50 & 1.97 & 2.49 & 2.46 & 2.74 & 2.19 & 2.48 & 2.19 & 2.53 & 2.18 & 2.57 \\
& DPARN & 2.75 & 2.97 & 2.35 & 2.69 & 2.27 & 2.69 & 2.34 & 2.75 & 2.57 & 2.79 & 2.30 & 2.59 & 2.36 & 2.66 & 2.33 & 2.66 \\
& ARN& \textbf{2.82} & \textbf{3.04} & \textbf{2.43} & \textbf{2.78} & \textbf{2.45} & \textbf{2.79} & \textbf{2.48} & \textbf{2.86} & \textbf{2.64} & \textbf{2.87} & \textbf{2.36} & \textbf{2.65} & \textbf{2.43} & \textbf{2.73} & \textbf{2.45} & \textbf{2.76} \\
\hline
\end{tabular}
\end{adjustbox}
\label{tbl:5}
\end{table*}

\begin{table*}[t]
\centering
\caption{Comparing causal ARN with other causal approaches to speech enhancement.}
\label{tbl_ablation}
\centering
\begin{adjustbox}{width=0.8\textwidth}
\begin{tabular}{|c|c|cc|cc|cc|cc|cc|cc|cc|cc|}
\hline
\multicolumn{2}{|c|}{Test Noise} & \multicolumn{8}{c|}{Babble} & \multicolumn{8}{c|}{Cafeteria} \\
\hline
\multicolumn{2}{|c|}{Test Corpus} & \multicolumn{2}{c|}{WSJ} & \multicolumn{2}{c|}{TIMIT} & \multicolumn{2}{c|}{IEEE Male} & \multicolumn{2}{c|}{IEEE Female} & \multicolumn{2}{c|}{WSJ} & \multicolumn{2}{c|}{TIMIT} & \multicolumn{2}{c|}{IEEE Male} & \multicolumn{2}{c|}{IEEE Female} \\
\hline
\multicolumn{2}{|c|}{Test SNR} & -5 dB & -2 dB & -5 dB & -2 dB & -5 dB & -2 dB & -5 dB & -2 dB & -5 dB & -2 dB & -5 dB & -2 dB & -5 dB & -2 dB & -5 dB & -2 dB \\
\hline
\hline
\multirow{8}{*}{ \rotatebox{90}{STOI (\%) }} & Mixture & 58.6 & 65.5 & 54.0 & 60.9 & 55.0 & 62.3 & 55.5 & 62.9 & 57.4 & 64.5 & 53.1 & 60.1 & 54.8 & 60.9 & 55.1 & 62.0 \\
& DCCRN & 79.0 & 86.7 & 69.6 & 79.6 & 66.2 & 79.3 & 67.2 & 80.9 & 78.6 & 85.7 & 71.6 & 80.2 & 68.9 & 78.0 & 73.4 & 83.4 \\
& RNN-IRM \cite{pandey2020cross} & 80.7 & 86.5 & 72.5 & 80.5 & 72.3 & 81.6 & 70.6 & 82.0 & 77.8 & 84.2 & 71.7 & 79.3 & 69.8 & 77.7 & 72.9 & 81.7 \\
& RNN-TCS \cite{pandey2020learning}  & 85.1 & 90.4 & 76.1 & 85.3 & 72.8 & 83.0 & 73.5 & 85.3 & 82.2 & 88.2 & 76.2 & 84.0 & 72.4 & 80.5 & 77.4 & 85.8 \\
& DCN & 83.7 & 89.2 & 73.0 & 82.3 & 69.6 & 80.7 & 69.6 & 82.7 & 81.3 & 87.3 & 74.5 & 82.6 & 70.5 & 79.3 & 74.6 & 84.1 \\
& DPARN & \textbf{88.5} & 92.3 & 79.6 & 87.4 & 75.3 & 84.9 & 79.0 & 88.7 & \textbf{85.1} & \textbf{90.0} & \textbf{79.0} & 85.9 & 74.6 & 82.5 & 79.8 & \textbf{87.8} \\
& ARN& 88.3 & \textbf{92.4} & \textbf{80.2} & \textbf{88.1} & \textbf{77.7} & \textbf{85.9} & \textbf{80.1} & \textbf{89.2} & 84.7 & \textbf{90.0} & \textbf{79.0} & \textbf{86.0} & \textbf{75.5} & \textbf{83.0} & \textbf{80.3} & \textbf{87.8} \\
\hline
\hline
\multirow{8}{*}{ \rotatebox{90}{PESQ }} & Mixture & 1.54 & 1.69 & 1.46 & 1.63 & 1.45 & 1.63 & 1.12 & 1.32 & 1.44 & 1.64 & 1.33 & 1.52 & 1.37 & 1.54 & 1.01 & 1.20 \\
& DCCRN & 2.14 & 2.47 & 1.82 & 2.21 & 1.74 & 2.19 & 1.56 & 2.11 & 2.19 & 2.50 & 1.99 & 2.27 & 1.94 & 2.28 & 1.98 & 2.36 \\
& RNN-IRM & 2.31 & 2.62 & 2.08 & 2.42 & 1.99 & 2.38 & 1.74 & 2.27 & 2.26 & 2.55 & 2.10 & 2.36 & 2.00 & 2.31 & 1.95 & 2.34 \\
& RNN-TCS & 2.32 & 2.63 & 2.00 & 2.36 & 2.00 & 2.39 & 1.83 & 2.34 & 2.22 & 2.50 & 2.03 & 2.30 & 2.04 & 2.36 & 2.06 & 2.42 \\
& DCN & 2.32 & 2.61 & 1.94 & 2.28 & 1.85 & 2.27 & 1.67 & 2.18 & 2.24 & 2.52 & 1.99 & 2.28 & 1.94 & 2.26 & 1.92 & 2.34 \\
& DPARN & \textbf{2.51} & 2.76 & 2.12 & 2.47 & 2.06 & 2.48 & 2.00 & 2.50 & \textbf{2.35} & 2.61 & \textbf{2.13} & \textbf{2.41} & 2.12 & 2.44 & 2.11 & 2.52 \\
& ARN & 2.50 & \textbf{2.78} & \textbf{2.15} & \textbf{2.52} & \textbf{2.16} & \textbf{2.53} & \textbf{2.10} & \textbf{2.56} & 2.34 & \textbf{2.62} & 2.12 & 2.39 & \textbf{2.14} & \textbf{2.45} & \textbf{2.17} & \textbf{2.51} \\
\hline
\end{tabular}
\end{adjustbox}
\label{tbl:5}
\end{table*}

\subsection{Comparison with Baselines}
Table I and Table II respectively report average STOI and PESQ scores over babble and cafeteria noises for causal and non-causal speech enhancement. First, we observe that DCCRN has the lowest objective scores for both causal and non-causal speech enhancement. This suggests that DCCRN is not effective in low SNR conditions, especially for the challenging IEEE corpus. Next, we observe that RNN-TCS is better than RNN-IRM for non-causal speech enhancement. For causal speech enhancement, RNN-TCS is better than RNN-IRM in terms of STOI, but for PESQ, RNN-IRM has similar or better scores for many test conditions.
Further, we notice that DCN does not obtain good scores on all the corpora. For many cases, DCN has even worse scores than RNN-IRM, suggesting that DCN fails to generalize to untrained corpora. Finally, we notice that even though DPARN scores are worse than ARN, the difference is less than $1\%$ for STOI and less than $0.1$ for PESQ in most of the cases. For some cases, such as non-causal enhancement for IEEE Male, DPARN is significantly worse than ARN.

In summary, using RNN with a smaller frame shift improves cross-corpus generalization. Complex spectral mapping and time-domain enhancement  are comparable to each other but better than ratio masking. Adding self-attention to RNN further improves cross-corpus generalization.  Although not comparable to ARN, DPARN obtains good cross-corpus generalization.

\subsection{Comparing Loss Functions}
We compare two loss functions, MSE and PCM. This comparison is to establish the importance of the PCM loss for high SNR enhancement. Results are given in Table III. We observe that, at $-5$ dB, PCM is better than MSE for WSJ and TIMIT, but similar to or worse than MSE for IEEE Male and Female. At $5$ dB PCM is better than MSE for all test conditions. Moreover, PESQ improvements are very significant for many cases. This suggests that, even though PCM is not consistently better in low SNR conditions, it is clearly a better loss function for high SNR speech enhancement. Therefore, we use the PCM loss for training models on VCTK and DNS challenge dataset, which require evaluation in relatively high SNR conditions.
\begin{table}[h]
\centering
\caption{Comparing MSE loss and PCM loss at SNRs of $-5$ dB and $5$ dB. a) Non-causal, b) causal.}
\label{tbl_ablation}
\centering
\begin{adjustbox}{width=0.48\textwidth}
\begin{tabular}{|c|c|c|c|cc|cc|cc|cc|}
\cline{4-12}
\multicolumn{1}{c}{} & \multicolumn{1}{c}{} & \multicolumn{1}{c|}{} & Test Corpus & \multicolumn{2}{c|}{WSJ} & \multicolumn{2}{c|}{TIMIT} & \multicolumn{2}{c|}{IEEE Male} & \multicolumn{2}{c|}{IEEE Female} \\
\cline{4-12}
\multicolumn{1}{c}{} & \multicolumn{1}{c}{} & \multicolumn{1}{c|}{} & Test SNR & -5 dB & 5 dB & -5 dB & 5 dB & -5 dB & 5 dB & -5 dB & 5 dB \\
\hline
\hline
\multirow{15}{*}{ \rotatebox{90}{Babble}} & \multirow{5}{*}{ \rotatebox{90}{STOI (\%) }} & & Noisy & 58.6 & 81.2 & 54.0 & 77.1 & 55.0 & 79.4 & 55.5 & 80.3 \\
\cline{3-12}
\ & & \multirow{2}{*}{ (a) } & MSE & 91.1 & 97.1 & 84.5 & 96.1 & \textbf{82.3} & 95.6 & \textbf{85.6} & 96.8 \\
& & & PCM & \textbf{92.0} & \textbf{97.6} & \textbf{84.9} & \textbf{96.7} & 81.5 & \textbf{95.9} & 84.8 & \textbf{97.2} \\
\cline{3-12}
& & \multirow{2}{*}{ (b) } & MSE & 88.3 & 96.6 & 80.2 & 95.3 & \textbf{77.7} & 94.6 & \textbf{80.1} & 96.1 \\
& & & PCM & \textbf{88.8} & \textbf{97.1} & \textbf{80.4} & \textbf{95.9} & \textbf{77.7} & \textbf{94.9} & 78.4 & \textbf{96.5} \\
\cline{2-12}
\cline{2-12}
& \multirow{5}{*}{ \rotatebox{90}{PESQ} } & & Noisy & 1.54 & 2.12 & 1.46 & 2.08 & 1.45 & 2.06 & 1.12 & 1.86 \\
\cline{3-12}
& & \multirow{2}{*}{ (a) } & MSE & 2.82 & 3.36 & 2.43 & 3.31 & \textbf{2.45} & 3.31 & \textbf{2.48} & 3.36 \\
& & & PCM & \textbf{2.98} & \textbf{3.57} & \textbf{2.52} & \textbf{3.48} & 2.41 & \textbf{3.40} & 2.47 & \textbf{3.57} \\
\cline{3-12}
& & \multirow{2}{*}{ (b) } & MSE & 2.50 & 3.25 & \textbf{2.15} & 3.13 & \textbf{2.16} & 3.14 & \textbf{2.10} & 3.18 \\
& & & PCM & \textbf{2.57} & \textbf{3.42} & 2.14 & \textbf{3.27} & 2.08 & \textbf{3.20} & 1.99 & \textbf{3.37} \\
\cline{2-12}
\cline{2-12}
& \multirow{5}{*}{ \rotatebox{90}{SI-SNR} } & & Noisy & -5.0 & 5.0 & -5.0 & 5.0 & -5.0 & 5.0 & -5.0 & 5.0 \\
\cline{3-12}
& & \multirow{2}{*}{ (a) } & MSE & \textbf{10.9} & \textbf{17.2} & \textbf{8.9} & \textbf{16.3} & \textbf{7.5} & \textbf{14.6} & \textbf{8.1} & 16.0 \\
& & & PCM & \textbf{10.9} & \textbf{17.2} & 8.8 & \textbf{16.3} & 6.7 & \textbf{14.6} & 7.5 & \textbf{16.1} \\
\cline{3-12}
& & \multirow{2}{*}{ (b) } & MSE & \textbf{9.1} & \textbf{16.6} & \textbf{7.2} & 15.5 & \textbf{5.3} & \textbf{13.9} & \textbf{5.8} & \textbf{15.3} \\
& & & PCM & \textbf{9.1} & \textbf{16.6} & \textbf{7.2} & \textbf{15.6} & \textbf{5.3} & 13.8 & 5.1 & \textbf{15.3} \\
\hline
\hline
\multirow{15}{*}{ \rotatebox{90}{Cafeteria} } & \multirow{5}{*}{ \rotatebox{90}{STOI (\%)} } & & Noisy & 57.4 & 81.2 & 53.1 & 76.2 & 54.8 & 77.0 & 55.1 & 78.4 \\
\cline{3-12}
& & \multirow{2}{*}{ (a) } & MSE & 88.3 & 96.4 & 82.7 & 95.1 & 80.6 & 94.0 & 85.3 & 95.9 \\
& & & PCM & \textbf{89.0} & \textbf{96.9} & \textbf{84.0} & \textbf{95.8} & \textbf{81.0} & \textbf{94.4} & \textbf{86.1} & \textbf{96.4} \\
\cline{3-12}
& & \multirow{2}{*}{ (b) } & MSE & 84.7 & 95.7 & 79.0 & 94.0 & \textbf{75.5} & 92.6 & 80.3 & 95.1 \\
& & & PCM & \textbf{85.2} & \textbf{96.2} & \textbf{79.4} & \textbf{94.7} & 75.2 & \textbf{92.9} & \textbf{80.3} & \textbf{95.7} \\
\cline{2-12}
\cline{2-12}
& \multirow{5}{*}{ \rotatebox{90}{PESQ }} & & Noisy & 1.44 & 2.12 & 1.33 & 2.02 & 1.37 & 1.98 & 1.01 & 1.79 \\
\cline{3-12}
& & \multirow{2}{*}{ (a) } & MSE & 2.64 & 3.26 & 2.36 & 3.20 & \textbf{2.43} & 3.22 & 2.45 & 3.20 \\
& & & PCM & \textbf{2.76} & \textbf{3.47} & \textbf{2.50} & \textbf{3.36} & \textbf{2.43} & \textbf{3.29} & \textbf{2.56} & \textbf{3.44} \\
\cline{3-12}
& & \multirow{2}{*}{ (b) } & MSE & 2.34 & 3.14 & \textbf{2.12} & 2.99 & \textbf{2.14} & 3.02 & \textbf{2.17} & 3.03 \\
& & & PCM & \textbf{2.38} & \textbf{3.31} & \textbf{2.12} & \textbf{3.13} & 2.02 & \textbf{3.09} & 2.16 & \textbf{3.24} \\
\cline{2-12}
\cline{2-12}
& \multirow{5}{*}{ \rotatebox{90}{SI-SNR }} & & Noisy & -5.0 & 5.0 & -5.0 & 5.0 & -5.0 & 5.0 & -5.0 & 5.0 \\
\cline{3-12}
& & \multirow{2}{*}{ (a) } & MSE & \textbf{9.6} & \textbf{16.2} & \textbf{9.0} & 15.9 & \textbf{7.2} & 14.2 & \textbf{8.4} & \textbf{15.9} \\
& & & PCM & 9.5 & 16.1 & \textbf{9.0} & \textbf{16.0} & 7.1 & \textbf{14.3} & \textbf{8.4} & \textbf{15.9} \\
\cline{3-12}
& & \multirow{2}{*}{ (b) } & MSE & 8.1 & \textbf{15.6} & 7.7 & \textbf{15.3} & \textbf{5.8} & \textbf{13.5} & \textbf{6.9} & 15.4 \\
& & & PCM & \textbf{8.2} & \textbf{15.6} & \textbf{7.8} & \textbf{15.3} & 5.6 & \textbf{13.5} & 6.8 & \textbf{15.5} \\
\hline
\end{tabular}
\end{adjustbox}
\label{tbl:5}
\end{table}

\subsection{Evaluation on VCTK}
We compare ARN trained on VCTK with baseline models in Table IV. We can see that non-causal ARN is significantly better than existing non-causal models. Causal ARN also obtains state-of-the-art results. However, the difference between second-best causal model and causal ARN is not as significant as between the second-best non-causal model and non-causal ARN.
\begin{table}[h]
\centering
\caption{Comparing ARN with baseline models on the VCTK dataset.}
\label{tbl_ablation}
\centering
\begin{adjustbox}{width=0.44\textwidth}
\begin{tabular}{|c|c|c|c|c|c|c|}
\cline{2-7}
\multicolumn{1}{c|}{} & PESQ & STOI (\%) & CSIG & CBAK & COVL & Causal? \\
\hline
Noisy & 1.97 & 91.5 & 3.35 & 2.44 & 2.63 & - \\
\hline
SEGAN \cite{pascual2017segan} & 2.16 & - & 3.48 & 2.94 & 2.80 & \xmark  \\
Wave U-Net \cite{macartney2018improved} & 2.4 & - & 3.52 & 3.24 & 2.96 & \xmark  \\
SEGAN-D \cite{phan2020improving} & 2.39 & - & 3.46 & 3.11 & 3.50 & \xmark  \\
MMSE-GAN \cite{soni2018time} & 2.53 & 93 & 3.80 & 3.12 & 3.14 & \xmark  \\
Metric-GAN \cite{fu2019metricgan}& 2.86 & - & 3.99 & 3.18 & 3.42 & \xmark \\
Metric-GAN+ \cite{fu2021metricgan+} & 3.15 & - & 4.14 & 3.16 & 3.64 & \xmark \\
DeepMMSE \cite{zhang2020deepmmse} & 2.95 & 94 & 4.28 & 3.46 & 3.64 & \xmark  \\
Koizumi 2020 \cite{koizumi2020speech} & 2.99 & - & 4.15 & 3.42 & 3.57 & \xmark  \\
HiFi-GAN \cite{su2020hifi} & 2.84 & - & 4.18 & 2.55 & 3.51 & \xmark  \\
T-GSA \cite{kim2020t} & 3.06 & - & 4.18 & 3.59 & 3.62 & \xmark  \\
DEMUCS \cite{defossez2020real} & 3.07 & 95 & 4.31 & 3.40 & 3.63 & \xmark \\
\hline
\textbf{NC-ARN} & \textbf{3.21} & \textbf{96 (95.7)} & \textbf{4.42} & \textbf{3.63} & \textbf{3.83} & \xmark  \\
\hline
\hline
Wiener & 2.22 & 93 & 3.23 & 2.68 & 2.67 & \checkmark \\
Deep Feature Loss  \cite{germain2018speech} & - & - & 3.86 & 3.33 & 3.22 & \checkmark \\
DeepMMSE \cite{zhang2020deepmmse} & 2.77 & 93 & 4.14 & 3.32 & 3.46 & \checkmark \\
MHANet \cite{nicolson2020masked} & 2.88 & 94 (93.6) & 4.17 & 3.37 & 3.53 & \checkmark \\
DEMUCS \cite{defossez2020real} & \textbf{2.93} & \textbf{95} & \textbf{4.22} & 3.25 & 3.52 & \checkmark \\
\hline
\textbf{ARN} & \textbf{2.96} & \textbf{95 (95.0)} & \textbf{4.21} & \textbf{3.46} & \textbf{3.59} & \checkmark \\
\hline
\end{tabular}
\end{adjustbox}
\label{tbl:5}
\end{table}

\subsection{Evaluation on Real Recordings}
Finally, we evaluate the causal ARN trained on the DNS challenge dataset on the blind test set of the second DNS challenge, which consists of $650$ real recordings and $50$ synthetic mixtures. Results are compared with a baseline NSNet in Table V. We observe that ARN is substantially better than NSNet for all the metrics and for all the speech classes. We also observe that ARN is able to improve both DNSMOS-SIG and DNSMOS-BAK for English, tonal, non-English and singing. For emotional speech, however, we see a good improvement in DNSMOS-BAK but reduction in DNSMOS-SIG. Overall, DNSMOS-OVR is substantially improved for all speech classes except for a slight reduction in singing and emotional speech. This may be due to very few training utterances for these two classes. Out of $347$K training utterances, only $5$K belong to the emotional class and only $2$K to the singing class.   

\begin{table}[h]
\centering
\caption{Evaluating ARN on the blind test set of the second DNS challenge using DNSMOS P.835.}
\label{tbl_ablation}
\centering
\begin{adjustbox}{width=0.44\textwidth}
\begin{tabular}{|c|c|c|c|c|c|c|c|}
\cline{3-8}
\multicolumn{1}{c}{}& \multicolumn{1}{c|}{}& Singing & Tonal & Non-English & English & Emotional & Overall \\
\hline
\multirow{3}{*}{ \rotatebox{90}{BAK }} & Noisy & 3.29 & 3.71 & 3.58 & 3.21 & 2.65 & 3.28 \\
& NSNet & 3.53 & 4.33 & 4.32 & 4.23 & 3.45 & 4.10 \\
& ARN & \textbf{3.70} & \textbf{4.39} & \textbf{4.45} & \textbf{4.46} & \textbf{3.54} & \textbf{4.27} \\
\hline
\multirow{3}{*}{ \rotatebox{90}{SIG} } & Noisy & 3.45 & 3.98 & 3.99 & 3.98 & \textbf{3.46} & 3.87 \\
& NSNet & 2.99 & 3.87 & 3.84 & 3.79 & 2.94 & 3.63 \\
& ARN & \textbf{3.54} & \textbf{4.07} & \textbf{4.18} & \textbf{4.22} & 3.00 & \textbf{3.98} \\
\hline
\multirow{3}{*}{\rotatebox{90}{ OVR }} & Noisy & \textbf{3.01} & 3.42 & 3.40 & 3.27 & \textbf{2.75} & 3.23 \\
& NSNet & 2.64 & 3.63 & 3.59 & 3.49 & 2.58 & 3.34 \\
& ARN& 2.99 & \textbf{3.78} & \textbf{3.88} & \textbf{3.92} & 2.61 & \textbf{3.65} \\
\hline
\end{tabular}
\end{adjustbox}
\label{tbl:5}
\end{table}

\section{Concluding Remarks}
In this study, we have proposed a novel ARN for time-domain speech enhancement to improve cross-corpus generalization. ARN comprises of RNN augmented with self-attention and feedforward blocks. We have trained ARN in a noise, speaker and corpus independent way and performed comprehensive evaluations on four untrained corpora for difficult nonstationary noises at low SNR conditions. Experimental results have demonstrated the superiority of ARN over competitive algorithms, such as RNN, DCCRN, DCN and DPARN.

We have found that RNN with a smaller frame shift, such as $4$ ms and $2$ ms, is an effective technique for speech enhancement with improved cross-corpus generalization. Further, we have revealed that although attention can obtain significant improvements, the types of attention mechanism do not make a big difference. We have also evaluated RNN and ARN for complex spectral mapping  and time-domain speech enhancement. A key finding is that complex spectral mapping and time-domain enhancement are similar to each other, but are significantly better than ratio masking when trained on a large corpus. Further, we have examined frame shifts of $4$ ms and $2$ ms and reported significantly better results with $2$ ms frame shift.

We have also trained ARN on the  VCTK dataset for speech quality improvement in relatively high SNR conditions and obtained state-of-the-art results. Additionally, we have trained a causal ARN to jointly perform dereverberation and denoising. For this training, we utilized speech, noises, and RIRs from the DNS challenge dataset. The evaluation on a blind test set using a non-intrusive quality metric demonstrates that ARN obtains strong quality improvements for real recordings. This illustrates that ARN is a highly effective and robust model for speech enhancement.

In the future, we plan to perform listening tests of ARN on IEEE utterances in low SNR conditions; IEEE sentences are widely used in speech intelligibility evaluations. Additionally, we plan to further investigate DPARN, as it is found to be effective for cross-corpus generalization. We have observed that the architectures with larger numbers of parameters, such as RNN and ARN, obtain better generalization compared to architectures with fewer parameters, such as convolutional neural networks. We plan to redesign DPARN architecture to expand its number of parameters to be comparable to that of RNN.   

Even though we have compared causal and non-causal approaches, we have not considered parameter efficiency and computational complexity of models, as the primary goal of this study is to improve cross-corpus generalization. ARN has significantly larger number of parameters compared to DPARN and DCN. A future research direction would be to optimize ARN for real-world applications by using techniques such as model compression and quantization \cite{tan2021towards}. A related research direction is to explore DNN architectures that have fewer number of parameters but provide good cross-corpus generalization.

\bibliographystyle{IEEEtran}
\bibliography{IEEEabrv, mybib}

\begin{thebibliography}{10}
\providecommand{\url}[1]{#1}
\csname url@samestyle\endcsname
\providecommand{\newblock}{\relax}
\providecommand{\bibinfo}[2]{#2}
\providecommand{\BIBentrySTDinterwordspacing}{\spaceskip=0pt\relax}
\providecommand{\BIBentryALTinterwordstretchfactor}{4}
\providecommand{\BIBentryALTinterwordspacing}{\spaceskip=\fontdimen2\font plus
\BIBentryALTinterwordstretchfactor\fontdimen3\font minus
  \fontdimen4\font\relax}
\providecommand{\BIBforeignlanguage}[2]{{%
\expandafter\ifx\csname l@#1\endcsname\relax
\typeout{** WARNING: IEEEtran.bst: No hyphenation pattern has been}%
\typeout{** loaded for the language `#1'. Using the pattern for}%
\typeout{** the default language instead.}%
\else
\language=\csname l@#1\endcsname
\fi
#2}}
\providecommand{\BIBdecl}{\relax}
\BIBdecl

\bibitem{loizou2013speech}
P.~C. Loizou, \emph{Speech Enhancement: Theory and Practice}, 2nd~ed.\hskip 1em
  plus 0.5em minus 0.4em\relax Boca Raton, FL, USA: CRC Press, 2013.

\bibitem{wang2017supervised}
D.~L. Wang and J.~Chen, ``Supervised speech separation based on deep learning:
  An overview,'' \emph{IEEE/ACM Transactions on Audio, Speech, and Language
  Processing}, vol.~26, pp. 1702--1726, 2018.

\bibitem{wang2014training}
Y.~Wang, A.~Narayanan, and D.~L. Wang, ``On training targets for supervised
  speech separation,'' \emph{IEEE/ACM Transactions on Audio, Speech and
  Language Processing}, vol.~22, pp. 1849--1858, 2014.

\bibitem{erdogan2015phase}
H.~Erdogan, J.~R. Hershey, S.~Watanabe, and J.~Le~Roux, ``Phase-sensitive and
  recognition-boosted speech separation using deep recurrent neural networks,''
  in \emph{ICASSP}, 2015, pp. 708--712.

\bibitem{lu2013speech}
X.~Lu, Y.~Tsao, S.~Matsuda, and C.~Hori, ``Speech enhancement based on deep
  denoising autoencoder.'' in \emph{INTERSPEECH}, 2013, pp. 436--440.

\bibitem{xu2015regression}
Y.~Xu, J.~Du, L.-R. Dai, and C.-H. Lee, ``A regression approach to speech
  enhancement based on deep neural networks,'' \emph{IEEE/ACM Transactions on
  Audio, Speech and Language Processing}, vol.~23, pp. 7--19, 2015.

\bibitem{weninger2015speech}
F.~Weninger, H.~Erdogan, S.~Watanabe, E.~Vincent, J.~Le~Roux, J.~R. Hershey,
  and B.~Schuller, ``Speech enhancement with {LSTM} recurrent neural networks
  and its application to noise-robust {ASR},'' in \emph{International
  Conference on Latent Variable Analysis and Signal Separation}, 2015, pp.
  91--99.

\bibitem{chen2016large}
J.~Chen, Y.~Wang, S.~E. Yoho, D.~L. Wang, and E.~W. Healy, ``Large-scale
  training to increase speech intelligibility for hearing-impaired listeners in
  novel noises,'' \emph{The Journal of the Acoustical Society of America}, vol.
  139, pp. 2604--2612, 2016.

\bibitem{fu2016snr}
S.-W. Fu, Y.~Tsao, and X.~Lu, ``{SNR}-aware convolutional neural network
  modeling for speech enhancement.'' in \emph{INTERSPEECH}, 2016, pp.
  3768--3772.

\bibitem{park2016fully}
S.~R. Park and J.~Lee, ``A fully convolutional neural network for speech
  enhancement,'' in \emph{INTERSPEECH}, 2017, pp. 1993--1997.

\bibitem{chen2017long}
J.~Chen and D.~L. Wang, ``Long short-term memory for speaker generalization in
  supervised speech separation,'' \emph{The Journal of the Acoustical Society
  of America}, vol. 141.

\bibitem{tan2018gated}
K.~Tan, J.~Chen, and D.~L. Wang, ``Gated residual networks with dilated
  convolutions for supervised speech separation,'' in \emph{ICASSP}, 2018, pp.
  21--25.

\bibitem{pandey2018adversarial}
A.~Pandey and D.~L. Wang, ``On adversarial training and loss functions for
  speech enhancement,'' in \emph{ICASSP}, 2018, pp. 5414--5418.

\bibitem{williamson2016complex}
D.~S. Williamson, Y.~Wang, and D.~L. Wang, ``Complex ratio masking for monaural
  speech separation,'' \emph{IEEE/ACM Transactions on Audio, Speech and
  Language Processing}, vol.~24, pp. 483--492, 2016.

\bibitem{paliwal2011importance}
K.~Paliwal, K.~W{\'o}jcicki, and B.~Shannon, ``The importance of phase in
  speech enhancement,'' \emph{Speech Communication}, vol.~53, pp. 465--494,
  2011.

\bibitem{choi2019phase}
H.-S. Choi, J.-H. Kim, J.~Huh, A.~Kim, J.-W. Ha, and K.~Lee, ``Phase-aware
  speech enhancement with deep complex {U-Net},'' in \emph{ICLR}, 2019.

\bibitem{hu2020dccrn}
Y.~Hu, Y.~Liu, S.~Lv, M.~Xing, S.~Zhang, Y.~Fu, J.~Wu, B.~Zhang, and L.~Xie,
  ``{DCCRN}: Deep complex convolution recurrent network for phase-aware speech
  enhancement,'' in \emph{INTERSPEECH}, 2020, pp. 2472--2476.

\bibitem{zhou2021complex}
L.~Zhou, Y.~Gao, Z.~Wang, J.~Li, and W.~Zhang, ``Complex spectral mapping with
  attention based convolution recurrent neural network for speech
  enhancement,'' \emph{arXiv preprint arXiv:2104.05267}, 2021.

\bibitem{fu2017complex}
S.-W. Fu, T.-y. Hu, Y.~Tsao, and X.~Lu, ``Complex spectrogram enhancement by
  convolutional neural network with multi-metrics learning,'' in \emph{Workshop
  on Machine Learning for Signal Processing}, 2017, pp. 1--6.

\bibitem{pandey2019exploring}
A.~Pandey and D.~L. Wang, ``Exploring deep complex networks for complex
  spectrogram enhancement,'' in \emph{ICASSP}, 2019, pp. 6885--6889.

\bibitem{tan2019learning}
K.~Tan and D.~L. Wang, ``Learning complex spectral mapping with gated
  convolutional recurrent networks for monaural speech enhancement,''
  \emph{IEEE/ACM Transactions on Audio, Speech, and Language Processing},
  vol.~28, pp. 380--390, 2019.

\bibitem{pandey2020learning}
A.~Pandey and D.~L. Wang, ``Learning complex spectral mapping for speech
  enhancement with improved cross-corpus generalization,'' in
  \emph{INTERSPEECH}, 2020, pp. 4511--4515.

\bibitem{fu2017raw}
S.-W. Fu, Y.~Tsao, X.~Lu, and H.~Kawai, ``Raw waveform-based speech enhancement
  by fully convolutional networks,'' \emph{arXiv:1703.02205}, 2017.

\bibitem{pascual2017segan}
S.~Pascual, A.~Bonafonte, and J.~Serrà, ``{SEGAN}: Speech enhancement
  generative adversarial network,'' in \emph{INTERSPEECH}, 2017, pp.
  3642--3646.

\bibitem{rethage2017wavenet}
D.~Rethage, J.~Pons, and X.~Serra, ``A wavenet for speech denoising,'' in
  \emph{ICASSP}, 2018, pp. 5069--5073.

\bibitem{qian2017speech}
K.~Qian, Y.~Zhang, S.~Chang, X.~Yang, D.~Flor{\^e}ncio, and
  M.~Hasegawa-Johnson, ``Speech enhancement using bayesian wavenet,'' in
  \emph{INTERSPEECH}, 2017, pp. 2013--2017.

\bibitem{fu2018end}
S.-W. Fu, T.-W. Wang, Y.~Tsao, X.~Lu, and H.~Kawai, ``End-to-end waveform
  utterance enhancement for direct evaluation metrics optimization by fully
  convolutional neural networks,'' \emph{IEEE/ACM Transactions on Audio,
  Speech, and Language Processing}, vol.~26, pp. 1570--1584, 2018.

\bibitem{pandey2019new}
A.~Pandey and D.~L. Wang, ``A new framework for {CNN}-based speech enhancement
  in the time domain,'' \emph{IEEE/ACM Transactions on Audio, Speech and
  Language Processing}, vol.~27, pp. 1179--1188, 2019.

\bibitem{pandey2019tcnn}
------, ``{TCNN}: Temporal convolutional neural network for real-time speech
  enhancement in the time domain,'' in \emph{ICASSP}, 2019, pp. 6875--6879.

\bibitem{pandey2020densely}
------, ``Densely connected neural network with dilated convolutions for
  real-time speech enhancement in the time domain,'' in \emph{ICASSP}, 2020,
  pp. 6629--6633.

\bibitem{giri2019attention}
R.~Giri, U.~Isik, and A.~Krishnaswamy, ``Attention wave-{U-Net} for speech
  enhancement,'' in \emph{WASPAA}, 2019, pp. 249--253.

\bibitem{pandey2021dense}
A.~Pandey and D.~L. Wang, ``Dense {CNN} with self-attention for time-domain
  speech enhancement,'' \emph{IEEE/ACM Transactions on Audio, Speech, and
  Language Processing}, vol.~29, pp. 1270--1279, 2021.

\bibitem{luo2020dual}
Y.~Luo, Z.~Chen, and T.~Yoshioka, ``{Dual-path RNN}: Efficient long sequence
  modeling for time-domain single-channel speech separation,'' in
  \emph{ICASSP}, 2020, pp. 46--50.

\bibitem{pandey2020dual}
A.~Pandey and D.~L. Wang, ``Dual-path self-attention {RNN} for real-time speech
  enhancement,'' \emph{arXiv:2010.12713}, 2020.

\bibitem{taal2011algorithm}
C.~H. Taal, R.~C. Hendriks, R.~Heusdens, and J.~Jensen, ``An algorithm for
  intelligibility prediction of time--frequency weighted noisy speech,''
  \emph{IEEE Transactions on Audio, Speech, and Language Processing}, vol.~19,
  pp. 2125--2136, 2011.

\bibitem{Pandey2018}
A.~Pandey and D.~L. Wang, ``A new framework for supervised speech enhancement
  in the time domain,'' in \emph{INTERSPEECH}, 2018, pp. 1136--1140.

\bibitem{pandey2020cross}
------, ``On cross-corpus generalization of deep learning based speech
  enhancement,'' \emph{IEEE/ACM Transactions on Audio, Speech, and Language
  Processing}, vol.~28, pp. 2489--2499, 2020.

\bibitem{vaswani2017attention}
A.~Vaswani, N.~Shazeer, N.~Parmar, J.~Uszkoreit, L.~Jones, A.~N. Gomez,
  {\L}.~Kaiser, and I.~Polosukhin, ``Attention is all you need,'' in
  \emph{NIPS}, 2017, pp. 5998--6008.

\bibitem{zhang2019self}
H.~Zhang, I.~Goodfellow, D.~Metaxas, and A.~Odena, ``Self-attention generative
  adversarial networks,'' in \emph{ICML}, 2019, pp. 7354--7363.

\bibitem{dong2018speech}
L.~Dong, S.~Xu, and B.~Xu, ``{Speech-Transformer}: a no-recurrence
  sequence-to-sequence model for speech recognition,'' in \emph{ICASSP}, 2018,
  pp. 5884--5888.

\bibitem{zhao2020monaural}
Y.~Zhao, D.~L. Wang, B.~Xu, and T.~Zhang, ``Monaural speech dereverberation
  using temporal convolutional networks with self attention,'' \emph{IEEE/ACM
  Transactions on Audio, Speech, and Language Processing}, vol.~28, pp.
  1598--1607, 2020.

\bibitem{kim2020t}
J.~Kim, M.~El-Khamy, and J.~Lee, ``{T-GSA}: Transformer with
  {G}aussian-weighted self-attention for speech enhancement,'' in
  \emph{ICASSP}, 2020, pp. 6649--6653.

\bibitem{koizumi2020speech}
Y.~Koizumi, K.~Yaiabe, M.~Delcroix, Y.~Maxuxama, and D.~Takeuchi, ``Speech
  enhancement using self-adaptation and multi-head self-attention,'' in
  \emph{ICASSP}, 2020, pp. 181--185.

\bibitem{nicolson2020masked}
A.~Nicolson and K.~K. Paliwal, ``Masked multi-head self-attention for causal
  speech enhancement,'' \emph{Speech Communication}, vol. 125, pp. 80--96,
  2020.

\bibitem{roy2021deeplpc}
S.~K. Roy, A.~Nicolson, and K.~K. Paliwal, ``{DeepLPC-MHANet: M}ulti-head
  self-attention for augmented {K}alman filter-based speech enhancement,''
  \emph{IEEE Access}, vol.~9, pp. 70\,516--70\,530, 2021.

\bibitem{ronneberger2015u}
O.~Ronneberger, P.~Fischer, and T.~Brox, ``{U-Net}: Convolutional networks for
  biomedical image segmentation,'' in \emph{International Conference on Medical
  Image Computing and Computer-assisted Intervention}, 2015, pp. 234--241.

\bibitem{merity2019single}
S.~Merity, ``Single headed attention {RNN}: Stop thinking with your head,''
  \emph{arXiv preprint arXiv:1911.11423}, 2019.

\bibitem{ba2016layer}
J.~L. Ba, J.~R. Kiros, and G.~E. Hinton, ``Layer normalization,''
  \emph{arXiv:1607.06450}, 2016.

\bibitem{ioffe2015batch}
S.~Ioffe and C.~Szegedy, ``Batch normalization: Accelerating deep network
  training by reducing internal covariate shift,'' in \emph{ICML}, 2015, pp.
  448--456.

\bibitem{hendrycks2016gaussian}
D.~Hendrycks and K.~Gimpel, ``Gaussian error linear units ({GELU}s),''
  \emph{arXiv:1606.08415}, 2016.

\bibitem{srivastava2014dropout}
N.~Srivastava, G.~Hinton, A.~Krizhevsky, I.~Sutskever, and R.~Salakhutdinov,
  ``Dropout: a simple way to prevent neural networks from overfitting,''
  \emph{The Journal of Machine Learning Research}, vol.~15, no.~1, pp.
  1929--1958, 2014.

\bibitem{panayotov2015librispeech}
V.~Panayotov, G.~Chen, D.~Povey, and S.~Khudanpur, ``Librispeech: an {ASR}
  corpus based on public domain audio books,'' in \emph{ICASSP}, 2015, pp.
  5206--5210.

\bibitem{paul1992design}
D.~B. Paul and J.~M. Baker, ``The design for the wall street journal-based
  {CSR} corpus,'' in \emph{Workshop on Speech and Natural Language}, 1992, pp.
  357--362.

\bibitem{garofolo1993darpa}
J.~S. Garofolo, L.~F. Lamel, W.~M. Fisher, J.~G. Fiscus, and D.~S. Pallett,
  ``{DARPA TIMIT} acoustic-phonetic continous speech corpus {CD-ROM}. {NIST}
  speech disc 1-1.1,'' \emph{NASA STI/Recon technical report n}, vol.~93, 1993.

\bibitem{rothauser1969ieee}
IEEE, ``{IEEE} recommended practice for speech quality measurements,''
  \emph{IEEE Transactions on Audio and Electroacoustics}, vol.~17, pp.
  225--246, 1969.

\bibitem{varga1993assessment}
A.~Varga and H.~J. Steeneken, ``Assessment for automatic speech recognition:
  {II. NOISEX-92}: A database and an experiment to study the effect of additive
  noise on speech recognition systems,'' \emph{Speech communication}, vol.~12,
  no.~3, pp. 247--251, 1993.

\bibitem{valentini2016investigating}
C.~Valentini-Botinhao, X.~Wang, S.~Takaki, and J.~Yamagishi, ``Investigating
  {RNN-based} speech enhancement methods for noise-robust text-to-speech.'' in
  \emph{SSW}, 2016, pp. 146--152.

\bibitem{defossez2020real}
A.~Defossez, G.~Synnaeve, and Y.~Adi, ``Real time speech enhancement in the
  waveform domain,'' in \emph{INTERSPEECH}, 2020, pp. 3291--3295.

\bibitem{reddy2021icassp}
C.~K. Reddy, H.~Dubey, V.~Gopal, R.~Cutler, S.~Braun, H.~Gamper, R.~Aichner,
  and S.~Srinivasan, ``{ICASSP} 2021 deep noise suppression challenge,'' in
  \emph{ICASSP}, 2021, pp. 6623--6627.

\bibitem{reddy2021interspeech}
C.~K. Reddy, H.~Dubey, K.~Koishida, A.~Nair, V.~Gopal, R.~Cutler, S.~Braun,
  H.~Gamper, R.~Aichner, and S.~Srinivasan, ``{INTERSPEECH }2021 deep noise
  suppression challenge,'' in \emph{INTERSPEECH}, pp. 2796--2800.

\bibitem{kingma2014adam}
D.~Kingma and J.~Ba, ``Adam: A method for stochastic optimization,'' in
  \emph{ICLR}, 2015.

\bibitem{paszke2017automatic}
A.~Paszke, S.~Gross, S.~Chintala, G.~Chanan, E.~Yang, Z.~DeVito, Z.~Lin,
  A.~Desmaison, L.~Antiga, and A.~Lerer, ``Automatic differentiation in
  {PyTorch},'' 2017.

\bibitem{micikevicius2018mixed}
P.~Micikevicius, S.~Narang, J.~Alben, G.~Diamos, E.~Elsen, D.~Garcia,
  B.~Ginsburg, M.~Houston, O.~Kuchaiev, G.~Venkatesh, and H.~Wu, ``Mixed
  precision training,'' in \emph{ICLR}, 2018.

\bibitem{reddy2020interspeech}
C.~K. Reddy, E.~Beyrami, H.~Dubey, V.~Gopal, R.~Cheng, R.~Cutler,
  S.~Matusevych, R.~Aichner, A.~Aazami, S.~Braun \emph{et~al.}, ``The
  {INTERSPEECH} 2020 deep noise suppression challenge: Datasets, subjective
  speech quality and testing framework,'' in \emph{INTERSPEECH}, 2020, pp.
  2492--2496.

\bibitem{rix2001perceptual}
A.~W. Rix, J.~G. Beerends, M.~P. Hollier, and A.~P. Hekstra, ``Perceptual
  evaluation of speech quality ({PESQ}) - a new method for speech quality
  assessment of telephone networks and codecs,'' in \emph{ICASSP}, 2001, pp.
  749--752.

\bibitem{reddy2021dnsmos}
C.~K. Reddy, V.~Gopal, and R.~Cutler, ``{DNSMOS P. 835: A} non-intrusive
  perceptual objective speech quality metric to evaluate noise suppressors,''
  \emph{arXiv preprint arXiv:2110.01763}, 2021.

\bibitem{macartney2018improved}
C.~Macartney and T.~Weyde, ``Improved speech enhancement with the
  {Wave-U-Net},'' \emph{arXiv:1811.11307}, 2018.

\bibitem{phan2020improving}
H.~Phan, I.~V. McLoughlin, L.~Pham, O.~Y. Ch{\'e}n, P.~Koch, M.~De~Vos, and
  A.~Mertins, ``Improving {GAN}s for speech enhancement,'' \emph{IEEE Signal
  Processing Letters}, vol.~27, pp. 1700--1704, 2020.

\bibitem{soni2018time}
M.~H. Soni, N.~Shah, and H.~A. Patil, ``Time-frequency masking-based speech
  enhancement using generative adversarial network,'' in \emph{ICASSP}, 2018,
  pp. 5039--5043.

\bibitem{fu2019metricgan}
S.-W. Fu, C.-F. Liao, Y.~Tsao, and S.-D. Lin, ``{MetricGAN: G}enerative
  adversarial networks based black-box metric scores optimization for speech
  enhancement,'' in \emph{ICML}, 2019, pp. 2031--2041.

\bibitem{fu2021metricgan+}
S.-W. Fu, C.~Yu, T.-A. Hsieh, P.~Plantinga, M.~Ravanelli, X.~Lu, and Y.~Tsao,
  ``{MetricGAN+: A}n improved version of metricgan for speech enhancement,'' in
  \emph{INTERSPEECH}, 2021, pp. 201--205.

\bibitem{zhang2020deepmmse}
Q.~Zhang, A.~Nicolson, M.~Wang, K.~K. Paliwal, and C.~Wang, ``{DeepMMSE: A}
  deep learning approach to {MMSE}-based noise power spectral density
  estimation,'' \emph{IEEE/ACM Transactions on Audio, Speech, and Language
  Processing}, vol.~28, pp. 1404--1415, 2020.

\bibitem{su2020hifi}
J.~Su, Z.~Jin, and A.~Finkelstein, ``{HiFi-GAN: H}igh-fidelity denoising and
  dereverberation based on speech deep features in adversarial networks,'' in
  \emph{INTERSPEECH}, 2020, pp. 4506--4510.

\bibitem{germain2018speech}
F.~G. Germain, Q.~Chen, and V.~Koltun, ``Speech denoising with deep feature
  losses,'' in \emph{INTERSPEECH}, 2019, pp. 2723--2727.

\bibitem{tan2021towards}
K.~Tan and D.~Wang, ``Towards model compression for deep learning based speech
  enhancement,'' \emph{IEEE/ACM Transactions on Audio, Speech, and Language
  Processing}, vol.~29, pp. 1785--1794, 2021.

\end{thebibliography}
\end{document}